\documentclass[aps,prd,reprint,longbibliography,nofootinbib,superscriptaddress]{revtex4-2}

\makeatletter\def\frontmatter@affiliationfont{\it\footnotesize}\makeatother

\usepackage{lmodern}
\usepackage[T1]{fontenc}
\usepackage{amsmath}
\usepackage{amssymb}

\usepackage{url}
\usepackage{multirow}
\usepackage{caption}
\usepackage[labelformat=simple]{subcaption}

\usepackage[nodayofweek]{datetime}
\usepackage{enumerate}
\usepackage{xspace}
\usepackage{xcolor}
\usepackage{graphicx}
\usepackage[pdftex,hidelinks]{hyperref}
\usepackage{bookmark}
\usepackage[compat=1.1.0]{tikz-feynman}
\usepackage{placeins}

\usepackage[capitalise]{cleveref}

\usepackage{import}
\usepackage{tabularx, booktabs}
\newcolumntype{Y}{>{\centering\arraybackslash}X}

\usepackage{esvect}

\usepackage{siunitx}

\usepackage[nomessages]{fp}
\newcommand{\wrapentry}[3]{\begin{minipage}[t]{\dimexpr#1}\setlength{\parindent}{0pt}\centering#2\end{minipage}}

\usepackage{xparse}   
\NewDocumentCommand{\rot}{O{45} O{1em} m}{\makebox[#2][l]{\rotatebox{#1}{#3}}}%

\newcommand{\px}{\ensuremath{{p}_x}\xspace}
\newcommand{\py}{\ensuremath{{p}_y}\xspace}
\newcommand{\pz}{\ensuremath{{p}_z}\xspace}
\newcommand{\pt}{\ensuremath{{p}_\mathrm{T}}\xspace}
\newcommand{\pxmiss}{\ensuremath{\vv{p}_\mathrm{x}^\text{miss}}\xspace}
\newcommand{\pymiss}{\ensuremath{\vv{p}_\mathrm{y}^\text{miss}}\xspace}
\newcommand{\ptmiss}{\ensuremath{\vv{p}_\mathrm{T}^\text{miss}}\xspace}
\newcommand{\ptvv}{\ensuremath{\vv{p}_\mathrm{T}^{\nu\bar\nu}}\xspace}
\newcommand{\ttbar}{\ensuremath{t\bar{t}}\xspace}
\newcommand{\vflows}{\mbox{\ensuremath{\nu}-Flows}\xspace}
\newcommand{\vvflows}{\mbox{\ensuremath{\nu^2}-Flows}\xspace}
\newcommand{\vvflowsPy}{\mbox{\ensuremath{\nu^2}-Flows~(Pythia8)}\xspace}
\newcommand{\vvflowspy}{\mbox{\ensuremath{\nu^2}-Flows~(Pythia8)}\xspace}
\newcommand{\mttbar}{\ensuremath{m_{\ttbar}}\xspace}
\newcommand{\dphill}{\ensuremath{\Delta\phi(\ell^+\ell^-)}\xspace}
\newcommand{\pttop}{\ensuremath{p_\mathrm{T}^{t}}\xspace}
\newcommand{\pttt}{\ensuremath{p_\mathrm{T}^{\ttbar}}\xspace}
\newcommand{\ytt}{\ensuremath{y_{\ttbar}}\xspace}
\newcommand{\vtruth}{\mbox{\ensuremath{\nu}-Truth}\xspace}
\newcommand{\vweight}{\mbox{\ensuremath{\nu}-Weighting}\xspace}
\newcommand{\ellipse}{\mbox{Ellipse}\xspace}
\newcommand{\nvidia}{\mbox{NVIDIA\textsuperscript{\textregistered}}\xspace}
\newcommand{\Wboson}{\ensuremath{W}\;boson\xspace}
\newcommand{\Wbosons}{\ensuremath{W}\;bosons\xspace}

\begin{document}
  \title{\texorpdfstring{$\nu^2$-Flows}{v2-Flows}: Fast and improved neutrino reconstruction in multi-neutrino final states with conditional normalizing flows}
  \author{John Andrew Raine}
  \email{john.raine@unige.ch}
  \affiliation{Département de physique nucléaire et corpusculaire, University of Geneva, Switzerland}
  \author{Matthew Leigh}
  \email{matthew.leigh@unige.ch}
  \affiliation{Département de physique nucléaire et corpusculaire, University of Geneva, Switzerland}
  \author{Knut Zoch}
  \affiliation{Département de physique nucléaire et corpusculaire, University of Geneva, Switzerland}
  \affiliation{Laboratory for Particle Physics and Cosmology, Harvard University, Cambridge, 02138 MA, USA}
  \author{Tobias Golling}
  \affiliation{Département de physique nucléaire et corpusculaire, University of Geneva, Switzerland}

  \begin{abstract}
      In this work we introduce \vvflows, an extension of the \vflows method to final states containing multiple neutrinos.
      The architecture can natively scale for all combinations of object types and multiplicities in the final state for any desired neutrino multiplicities.
      In \ttbar dilepton events, the momenta of both neutrinos and correlations between them are reconstructed more accurately than when using the most popular standard analytical techniques, and solutions are found for all events.
      Inference time is significantly faster than competing methods, and can be reduced further by evaluating in parallel on graphics processing units.
      We apply \vvflows to \ttbar dilepton events and show that the per-bin uncertainties in unfolded distributions is much closer to the limit of performance set by perfect neutrino reconstruction than standard techniques.
      For the chosen double differential observables \vvflows results in improved statistical precision for each bin by a factor of 1.5 to 2 in comparison to the Neutrino Weighting method and up to a factor of four in comparison to the Ellipse approach.
  \end{abstract}

  \maketitle

  \section{Introduction}
  At collider experiments in particle physics, such as those at the Large Hadron Collider~(LHC)~\cite{LHC}, beams of hadrons or leptons are accelerated to high energies and collided together.
These collisions result in an array of particles which are studied by experiments comprising complex detectors surrounding the interaction points.
General purpose detectors, such as ATLAS~\cite{ATLAS} and CMS~\cite{CMS}, are designed to record and reconstruct nearly all stable particles predicted in the standard model of particle physics~(SM).
From these reconstructed stable particles, both precision measurements of the SM as well as searches for new phenomena beyond the SM~(BSM) are performed.

Neutrinos, stable particles produced in many collisions, interact only through the electroweak force and traverse the detectors without leaving a trace.
Their presence in collisions is inferred from a momentum imbalance in the transverse plane perpendicular to the beam axis.
This imbalance, known as the missing transverse momentum \ptmiss, is calculated from the negative vector sum of the transverse momenta of all reconstructed objects in the transverse plane.

In order to reconstruct individual neutrinos, and thus fully reconstruct a single event, underlying assumptions need to be made on their origin and multiplicity.
The \ptmiss serves as a proxy for all unobserved particles in the collision, but doesn't indicate how many were present, or how the momentum should be shared in the case of multiple neutrinos.
Furthermore, at hadron colliders there is no experimental equivalent for the missing longitudinal momentum.

Several approaches are used to reconstruct neutrinos using \ptmiss and by setting constraints on the invariant masses of intermediate particles.
In the case of top quark pair production (\ttbar) in the semileptonic decay channel, where there is one lepton and one neutrino in the event, the neutrino momentum is typically reconstructed by solving the longitudinal momentum component $\pz$ under the assumption that the invariant mass of the lepton-neutrino pair is exactly that of the \Wboson~\cite{QuadraticConvention, QuadraticConvention2, ATLAS:2019hxz, CMS:2021vhb, ATLAS:2015pfy, CMS:2018quc, ATLAS:2018fwq, ATLAS:2019guf, ATLAS:2022waa}.
For the case of events with two neutrinos, such as \ttbar production in the dileptonic channel, more complicated methods~\cite{NuW,Sonnenschein:2005ed,ellipse} are employed in order to resolve the share of momentum between the two neutrinos in the event~\cite{CMS:2011acs, CMS:2012tdr, ATLAS:2013gil, CMS:2014rdf, ATLAS:2014aus, ATLAS:2015ysm, D0:2015dxa, ATLAS:2015ysm, CMS:2016piu, CMS:2016ypc, ATLAS:2016bac, ATLAS:2016pbv, ATLAS:2019zrq, CMS:2019nrx, ATLAS:2022waa}.
These approaches still require the invariant mass of the two lepton-neutrino pairs to match the \Wboson mass, and in addition require the invariant mass of the lepton-neutrino-jet triplets to match the top-quark invariant mass.

In \vflows~\cite{vflows} we use conditional normalizing flows~\cite{NormFlows1,NormFlows2,cINNs} to learn the probability distribution of the neutrino momentum vector given the observed objects in an event for semileptonic \ttbar events.
From the learned conditional probability distribution, solutions can be sampled per event and the most probable solution can be determined from the learned likelihood of the solution.
In this work, we extend \vflows to more challenging final states with multiple neutrinos.
We focus on \ttbar final states
in which both top quarks decay semileptonically.
This results in an expected final state with exactly two oppositely charged leptons, at least two jets, of which two should originate from $b$-quarks, and two neutrinos.
In comparison to the single neutrino case, where the main challenge arises in recovering the longitudinal momentum component of the neutrino, events with multiple neutrinos have the additional complexity of how to share the total missing momentum vector between all neutrinos in the final state.
We show that \vflows yields improved reconstruction performance in comparison to standard approaches and demonstrate the direct benefit to the statistical precision in a simplified double differential \ttbar dilepton analysis in observables such as the invariant mass of the \ttbar system, $m_{\ttbar}$, and the angular separation of the two leptons, $\dphill$.

The repository\footnote{\url{https://github.com/rodem-hep/nu2flows}} and data\footnote{\url{https://zenodo.org/record/8113516}~\cite{ZenodoData_dilep}} used in this work are both made publicly available.

  \section{Method}
  
\subsection{\texorpdfstring{\vvflows}{v2-Flows}}

\begin{figure*}[ht]
    \centering
    \includegraphics[width=0.75\textwidth]{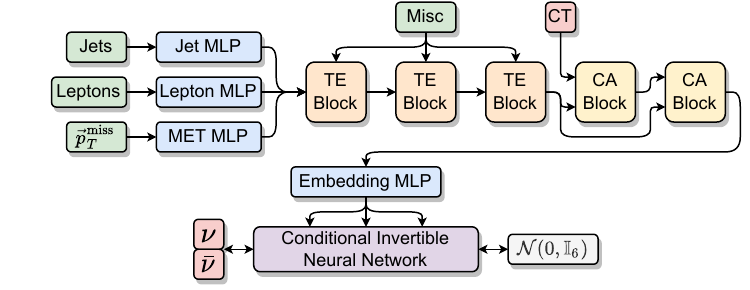}
    \caption{A schematic of the \vvflows network for learning the conditional likelihood of multiple neutrinos in the event.
    The network uses a transformer encoder (TE) with cross-attention (CA) with a learnable class token (CT) to embed an event representation for any multiplicity of physics objects.
    This operation is permutation invariant and can operate on any jet and lepton multiplicity.
    Each physics object has its own dedicated embedding network and additional event information (Misc) is used to condition the transformer encoder blocks.
    The representation vector is used to condition the transformation with the normalizing flow.}
    \label{fig:nunuflows}
\end{figure*}

In \vflows, conditional normalizing flows are used to capture the distribution of possible solutions for the neutrino momenta given the reconstructed momenta of observed objects in a collision.
The overall model comprises two components, the event feature extraction, and the conditional normalizing flow.
The event feature extraction learns a representative vector of the event from the collections of reconstructed objects, namely the jets and leptons, and \ptmiss.
The resulting vector is used as a conditional input to the normalizing flow, in order to learn the conditional density of all possible neutrino solutions from the training data.

In this work we extend the initial \vflows architecture to predict multiple neutrinos and accommodate any number of leptons in addition to jets by using attention transformers~\cite{vaswani2017attention}.
We label this architecture as \vvflows to distinguish it from the general method.
In order to handle two neutrinos we double the dimensionality of the conditional normalizing flow (from three to six).
The neutrinos are also always predicted in the same order for each event, with the momentum of the neutrino followed by the momentum of the anti-neutrino.
When increasing the neutrino multiplicity further, the same procedure is used together with a predefined ordering for the neutrinos.
The architecture of the normalizing flow is otherwise kept largely the same. 

The most substantial optimisation has been performed on the feature extraction network.
The feature extraction network attempts to produce a contextual vector, specific to each event, to guide the transformations within the normalizing flow.
In the single lepton case, \vflows uses an attention pooled deep set~\cite{DeepSets} to process the jets, with \pxmiss, \pymiss, the lepton four momentum, and some event level information as extra conditional information.
To extend \vvflows to multiple leptons, we require a permutation invariant architecture that can accommodate a variable number of both jets and leptons, motivating the move to attention transformers.

To train \vvflows all jets and leptons are represented by their four-momentum vectors in the form $(\px, \py, \pz, \log E)$.
Jets are assigned an additional binary decision on whether they are tagged as originating from a $b$-quark.
Leptons are identified as being either an electron or a muon, as well as whether they had positive or negative charge.
The target neutrino momenta are expressed as $(\px, \py, \pz)$ for both the neutrino and anti-neutrino.
The full set of inputs to the network are provided in \cref{tab:inputs}.
The coordinates chosen to describe the input and target object kinematics were optimised in a grid search.

\begin{table}[ht]
    \caption{The different input observables used as inputs to the feature extraction network.}
    \label{tab:inputs}
    \centering
    \resizebox{\columnwidth}{!}{
    \begin{tabular}{r c l}
    \toprule
    Category & Variables & Description\\
    \midrule
    \ptmiss & $p_x^\text{miss}$, $p_y^\text{miss}$ & Missing transverse momentum 2-vector \\ [2ex]
    \multirow{3}{*}{Leptons} & $p_x^{\ell}$, $p_y^{\ell}$, $p_z^\ell$, $\log E^\ell$ & Lepton momentum 4-vector \\
    & $q^\ell$ & Lepton charge\\
    & $\ell^{flav}$ &  Whether lepton is an electron or muon \\ [2ex]
    \multirow{2}{*}{Jets} & $p_x^{j}$, $p_y^j$, $p_z^j$,$\log E^j$ & Jet momentum 4-vector \\
    & $isB$ &  Whether jet passes $b$-tagging criteria \\ [2ex]
    Misc & $N_{\text{jets}}$, $N_{\text{bjets}}$ & Jet and $b$-jet multiplicities in the event\\
    \bottomrule
    \end{tabular}
    }
\end{table}

A schematic of the new architecture for \vvflows is shown in \cref{fig:nunuflows}, which makes use of attention transformers and object specific embedding networks.
Initially, the jets, leptons and \ptmiss are all independently embedded into higher dimensional space using object specific multi-layer perceptrons (MLP).%
\footnote{Other final state objects such as photons and tau leptons can also be accommodated by embedding them additional MLPs for each particle type.}
The embedded objects subsequently interact through a transformer encoder using several layers of multi-headed attention.
Additional event information (Misc) containing object multiplicities, is injected into the network as conditional information by concatenating the vector to each token within the transformer encoder blocks.
To obtain a single global vector from transformer as our conditioning vector for the normalizing flow, we perform repeated cross-attention with a learnable class token (CT), a common technique used in vision transformers \cite{VisionTransformers}.

The \vvflows model in this paper comprises three transformer encoder blocks and two cross-attention blocks, each with an embedding dimension of 128 and 16 attention heads.
All MLPs in the network have a single hidden layer with 256 neurons and use the LeakyReLU activation~\cite{LeakyRelu} and Layer-Normalization~\cite{LayerNorm}.
The output of the transformer is passed through an MLP to produce the context tensor for the flow with a dimension of 128.
The invertible neural network employs 10 rational quadratic spline~(RQS) coupling blocks ~\cite{NeuralSplines} interspersed with LU-decomposed linear layers, implemented with the \texttt{nflows} package~\cite{NFlows} and \texttt{Pytorch}~v2.0~\cite{Pytorch}.
Each RQS has 10 bins with linear tail bounds outside $\pm 4$.
The conditional normalizing flow is trained with the standard maximum likelihood estimation loss obtained through the change of variables formula, and transforms the input neutrino momenta to a standard multivariate normal distribution.
The entire \vvflows model, including the transformer, has around 600\,000 trainable parameters.

We train the \vvflows model for 100 epochs using the AdamW optimizer~\cite{AdamW} with a learning rate cycling from \num{e-8} to \num{e-3} and back every 50 epochs.
We use weight decay with a strength of \num{e-4}.

\subsection{Reference methods}

Several analytical techniques have been proposed to solve the reconstruction of the two neutrinos in dilepton \ttbar events.
Amongst these are \emph{Neutrino weighting}~\cite{NuW} (\vweight), an \emph{algebraic} solution~\cite{Sonnenschein:2005ed}, and the \emph{Ellipse} method~\cite{ellipse}.
These have been successfully employed in a wide range of measurements at the Tevatron and LHC~\cite{CMS:2011acs, CMS:2012tdr, ATLAS:2013gil, CMS:2014rdf, ATLAS:2014aus, ATLAS:2015ysm, D0:2015dxa, ATLAS:2015ysm, CMS:2016piu, CMS:2016ypc, ATLAS:2016bac, ATLAS:2016pbv, ATLAS:2019zrq, CMS:2019nrx, ATLAS:2022waa}, most notably \vweight.
In this work we compare \vvflows to the \vweight due to its common usage but also the \ellipse method due to its reduced computation time.


\subsubsection*{\texorpdfstring{\vweight}{Neutrino Weighting}}

In \vweight the kinematic properties of the neutrino are extracted from the identified leptons, jets and missing transverse momentum in the event.
For our implementation we follow the prescription described in Ref.~\cite{ATLAS:2019zrq}.
Constraints on neutrino solutions using the invariant mass of the top quark and \Wboson
\begin{gather}
    \label{eq:mass_constraints}
    (\ell_{1, 2} + \nu_{1, 2})^2 = m_w^2 = (\SI{80.38}{\giga\electronvolt})^2, \\
    (\ell_{1, 2} + \nu_{1, 2} + b_{1, 2})^2 = m_t^2 = (\SI{172.5}{\giga\electronvolt})^2,
\end{gather}
are applied, where $\ell_{1, 2}$, $\nu_{1, 2}$, and $b_{1, 2}$ represent the four-momenta of the charged leptons, neutrinos and $b$-tagged jets.
However, this is not enough to fully constrain the kinematics of the neutrinos.
Therefore, the neutrino and anti-neutrino rapidities ($\eta^\nu$ and $\eta^{\bar\nu}$) are individually hypothesized and tested.
For each pair of values for $\eta^\nu$ and $\eta^{\bar\nu}$, we solve the  mass equations in \cref{eq:mass_constraints}, yielding two possible solutions for the full pair of neutrino kinematics.
Each solution produces an inferred missing transverse momentum $\ptvv$ vector which can then be compared to the observed \ptmiss.
This comparison defines a weight
\begin{equation}
    \label{eq:neutrino_weight}
    w = \exp \left( - \frac{||\ptmiss - \ptvv||_2^2}{2 \sigma^2} \right),
\end{equation}
where $\sigma$ is a fixed resolution scale related to the \ptmiss reconstruction in the detector.
In \vweight the hypothesis that maximizes $w$ is chosen as the correct solution.
To find all solutions we perform a grid search of $\eta^\nu$ and $\eta^{\bar\nu}$ with values ranging from -5 to 5 with a step size of 0.2.
For each of these we also need to test all combinations of assigning $b$-tagged jets to each of the $b$-quarks from the \ttbar decay.
This represents a very costly computation scan that considers only discrete $\eta$ values.

Another significant drawback of this method is that despite the large number of neutrino solutions being tested, it is still possible that the constraint systems are not solvable.
This can be due to mis-measurement or mis-assignment of the leptons, jets, or \ptmiss.
Alternatively, this can arise from the masses of either the top quarks or \Wbosons in the event deviating from the nominal values.
Therefore, to increase the success rate of this method we also iterate over different values of $m_t$ from 171 to \SI{174}{\giga\electronvolt} with a step size of \SI{0.5}{\giga\electronvolt}.
This further increases the computational requirements, but increases the efficiency of finding a solution for each event.

After selecting the solution with the highest $w$, any solution which results in $m_{t \bar t} < \SI{300}{\giga\electronvolt}$ or where either of the two reconstructed top quarks have negative energy is rejected.
The \vweight method is unable to find a valid solution on the nominal dataset around 5\% of the time.

\subsubsection*{Ellipse method}

The \ellipse method is derived from a geometric approach to analytically constrain neutrino kinematics explicitly in processes where top quarks decay into leptons and neutrinos \cite{ellipse}.
For a single neutrino, its momentum can be calculated as a function of the 4-momenta of the $b$-quark and the charged lepton, the \Wboson mass, and the top quark mass.
The solution set for this function defines the surface of an ellipse.
By combining this information with the observed \ptmiss, the solution set collapses to a unique value.
For events with two neutrinos in the final state, the method is extended to  calculate the solution for neutrino pairs which are most likely to have produced the observed \ptmiss.

We use the implementation from the authors of the \ellipse method\footnote{Implementation available at \url{github.com/betchart/analytic-nu}}.
To solve the $b$-jet combinatorics we use a simple minimum $\Delta R$ matching between the leptons and the $b$-jets in the event.
To reduce the combinations we only take the two leading jets in \pt passing $b$-tagging criteria.
If the $\Delta R$ matching yields no solutions for the neutrino kinematics using the ellipse method, the opposite association is tested.

The drawbacks of this approach is that it requires accurate matching between each lepton and the associated $b$-jet in the event.
Furthermore, like \vweight, it requires one to make hard assumptions on the mass of the \Wboson and top quark.
Finally, it is possible that the method can yield no solutions, just like \vweight.
The implementation used in this work fails to find solutions in 22\% of \ttbar dilepton events.
In comparison to \vweight, the \ellipse method requires much less computational resources per event.

  \section{Dataset}
  In this work, \vvflows is applied to simulated \ttbar events where both top quarks decay semileptonically,
resulting in a final state with exactly two leptons ($\ell$)%
\footnote{Leptons is used to designate either electrons ($e$) or muons ($\mu$).}
, two neutrinos ($\nu$) and two jets initiated by $b$-hadrons ($b$-jets).
Additional jets arise from initial and final state radiation.
All events are simulated in proton-proton collisions at a centre-of-mass energy of \mbox{$\sqrt{s}=13$~TeV}.
Two different samples are generated, each using a different generator for the hard interactions in the matrix element. %

In the nominal sample, hard interactions are simulated using MadGraph5\_aMC@NLO~\cite{MadGraph}~(v3.1.0). 
The top-quark mass $m_t$ is set to 173~GeV for all events.
All events are interfaced to Pythia8~\cite{Pythia}~(v8.243) for the parton shower and hadronisation, using the \mbox{NNPDF2.3LO} PDF set~\cite{PartonDFs} with $\alpha_S(m_Z) = 0.130$ using the LHAPDF~\cite{PartonDAccessLHCC} framework.

In the alternative sample, both the hard interactions and parton shower are simulated with Pythia8~(v8.307) with the Monash tuned set of
parameters~\cite{Monash} at leading order accuracy.
The same PDF set is used as for the nominal sample.

The detector response is simulated using Delphes~\cite{Delphes}~(v3.4.2) with a parametrisation similar to the response of the ATLAS detector~\cite{ATLAS}.
Jets are reconstructed using the anti-$k_t$ clustering algorithm~\cite{AntiKt} with a radius parameter of $R=0.4$ using the FastJet package~\cite{FastJet}.
All jets are required to have a transverse momentum $\pt > 25$~GeV and fall within $|\eta| < 2.5$.
A $b$-tagging working point corresponding to 70\% inclusive signal efficiency is used to identify jets as originating from $b$-hadrons.
Up to 10 jets are selected per event, ordered in descending \pt.

Events are required to have at least two $b$-tagged jets, and two oppositely charged leptons, each with $\pt > 15$~GeV and $|\eta| < 2.5$.
Truth association of jets to the $b$-quarks in the \ttbar hard scatter is performed using a $\Delta R$ matching, with partons matched to jets within $\Delta R < 0.4$.
Events where multiple partons are matched to the same jet are removed.
Truth association of the lepton to the parent top quark is performed assuming there is no charge misidentification, and
the true neutrino momenta are taken directly from simulation.

In total there are 1.02 million events in the nominal sample and 1.4 million events in the alternative sample passing all selection requirements.
940,000 (970,000) events from the nominal (alternative) sample are used to train the network, with 80,000 nominal samples used for evaluation.


  \section{Results}
  \subsection{Neutrino reconstruction}
  The first measure of performance is to validate that \vvflows is able to correctly reconstruct the momenta and relative positions of the neutrino pair.
To establish a baseline for the impact of neutrino reconstruction in all distributions, we define \vtruth to be the case where the truth neutrinos are reconstructed perfectly, but all other objects in the event remain the same.
This defines the ground truth of the target distributions and is also the upper limit in performance for any neutrino reconstruction approach.
We compare the reconstruction performance with \vvflows to the \vweight and \ellipse methods.


\begin{figure*}[htpb]
    \centering
        \includegraphics[width=0.32\textwidth]{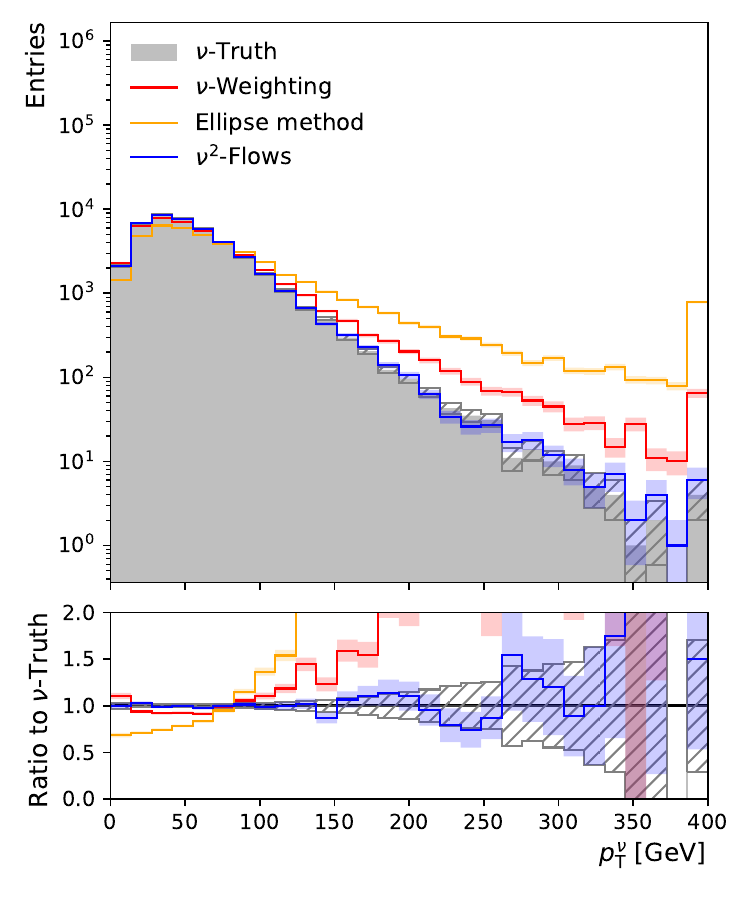}
        \includegraphics[width=0.32\textwidth]{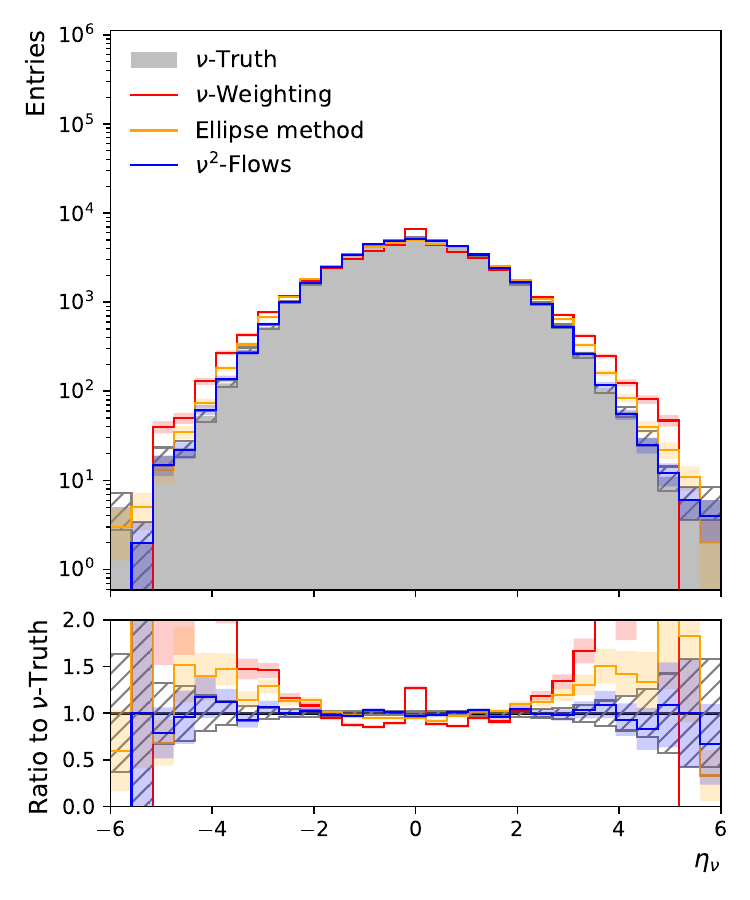}
        \includegraphics[width=0.32\textwidth]{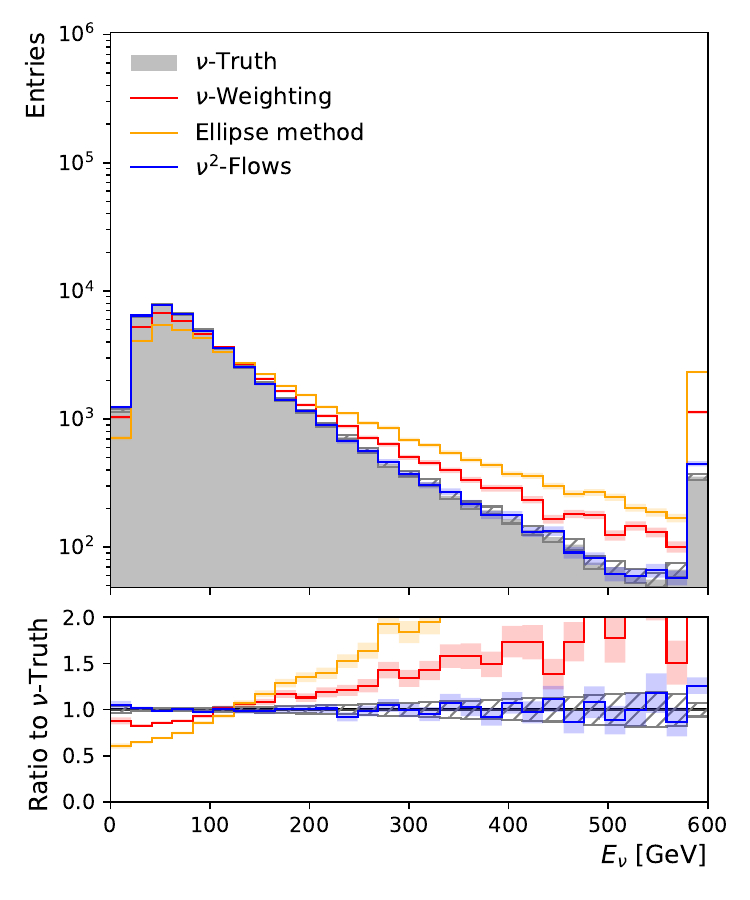}
    \caption{The kinematics of the reconstructed (anti-)neutrinos for the three reconstruction methods and \vtruth (shaded grey). The hashed areas represent statistical uncertainties in the \vtruth prediction.
    \label{fig:vvbarkinematics}
    }
\end{figure*}

\begin{figure*}[htpb]
    \centering
    \includegraphics[width=0.32\textwidth]{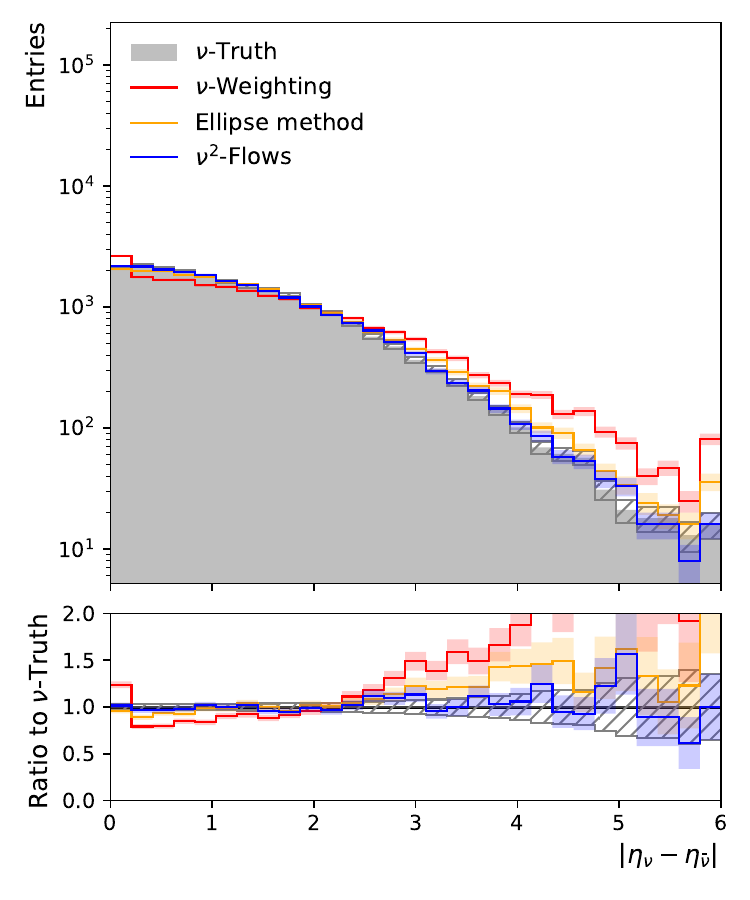}
    \includegraphics[width=0.32\textwidth]{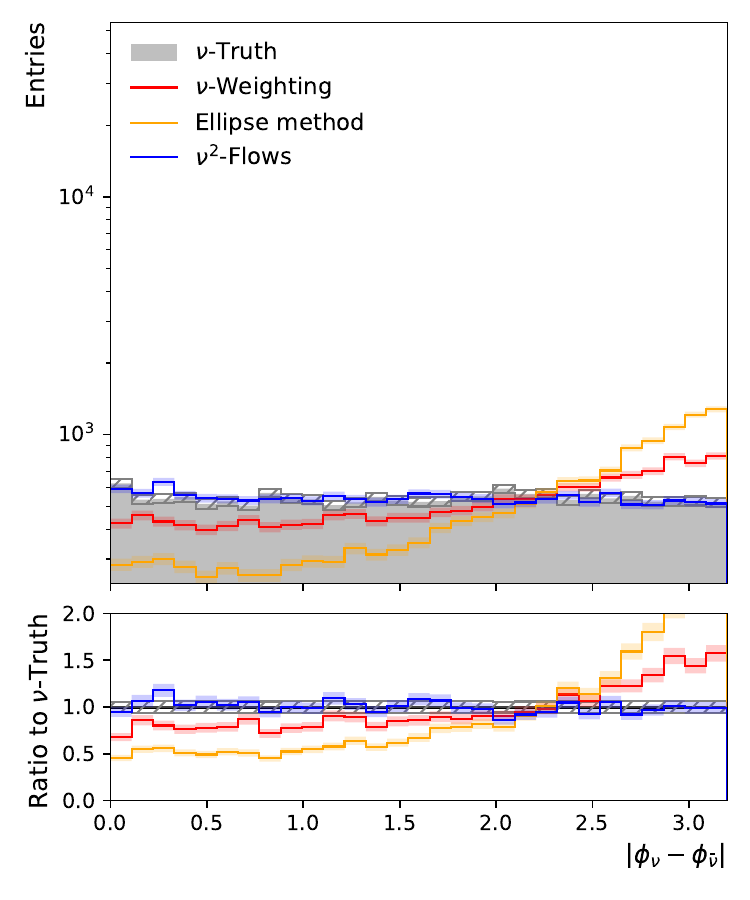}
    \includegraphics[width=0.32\textwidth]{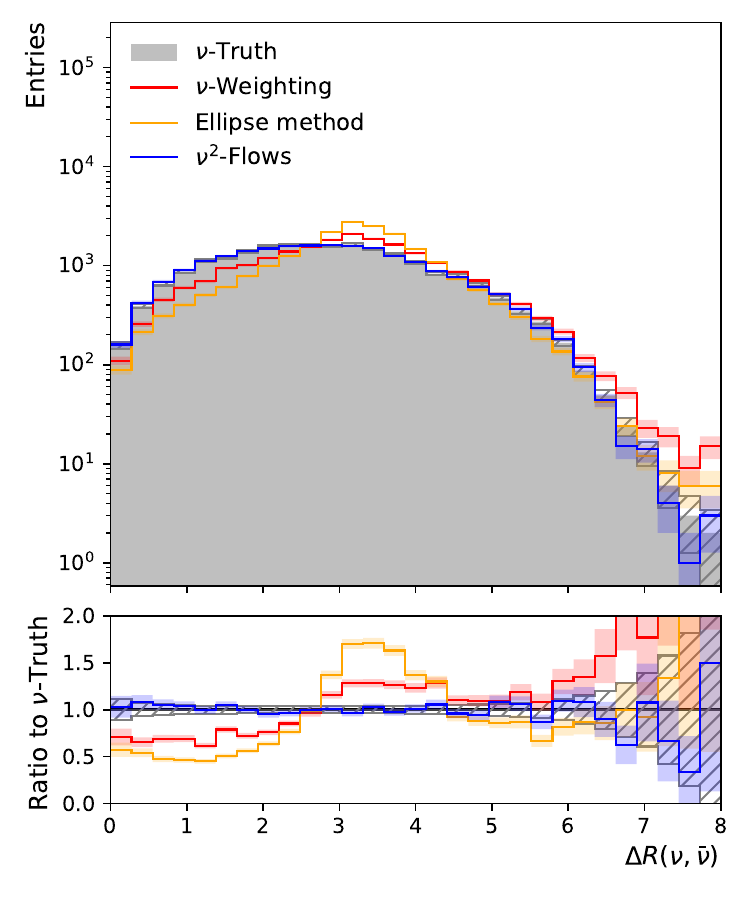}
    \caption{The angular separation in $\eta$ and $\phi$ between the reconstructed neutrino pair per event for the three reconstruction methods and \vtruth (shaded grey). The hashed areas represent statistical uncertainties in the \vtruth prediction.
    \label{fig:vreco_delta}
    }
\end{figure*}

The individual neutrino kinematics and the angular separation between the two neutrinos
are shown in \cref{fig:vvbarkinematics,fig:vreco_delta}.
Here we can see that both \vweight and \ellipse overestimate the neutrino transverse momenta and energy and tend to prefer central neutrinos with a visible peak at $\eta=0$.
The neutrino pair is also predominantly predicted to be back to back in $\phi$, with the opening angle between them showing a large degree of tension with the ground truth.
In all neutrino kinematic distributions, \vvflows is able to reproduce the ground truth accurately, with slight discrepancies only visible in the low statistic tails.

\begin{figure*}[htbp]
    \centering
    \includegraphics[width=0.32\textwidth]{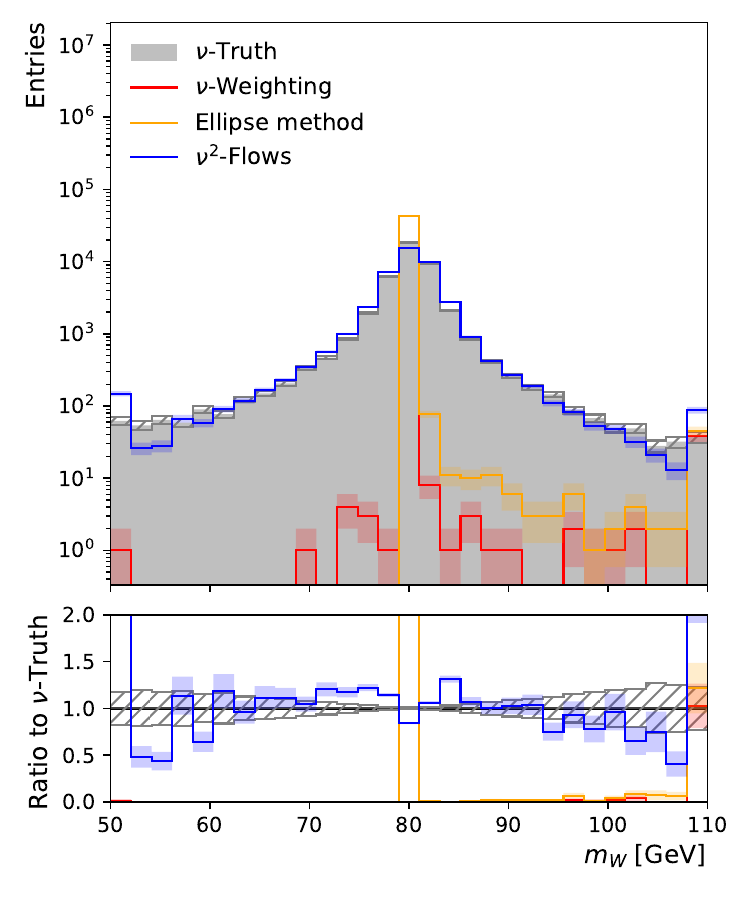}
    \includegraphics[width=0.32\textwidth]{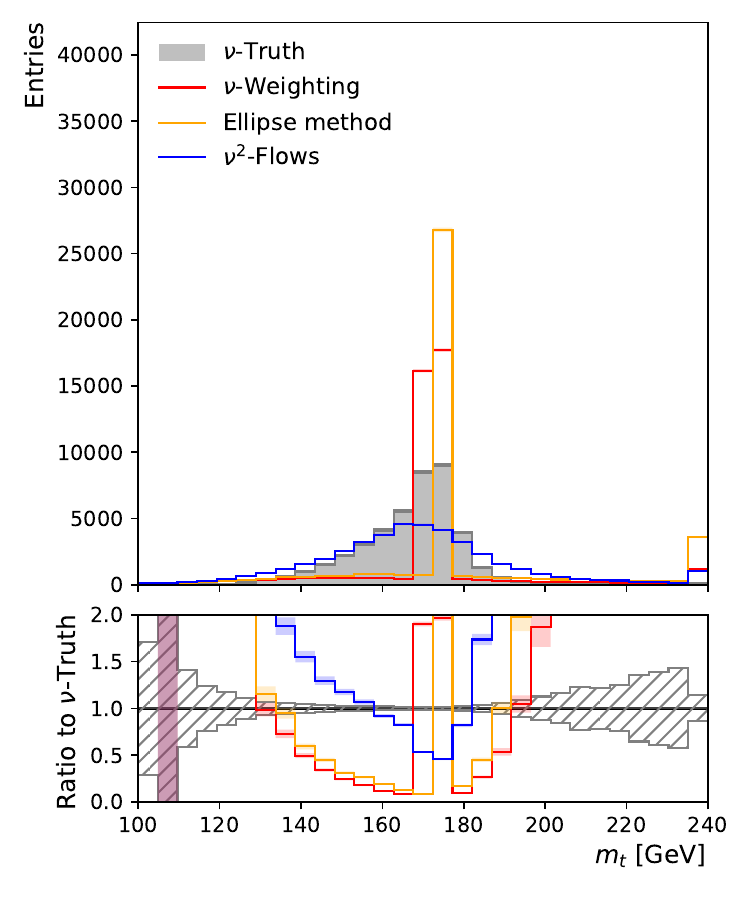}
    \includegraphics[width=0.32\textwidth]{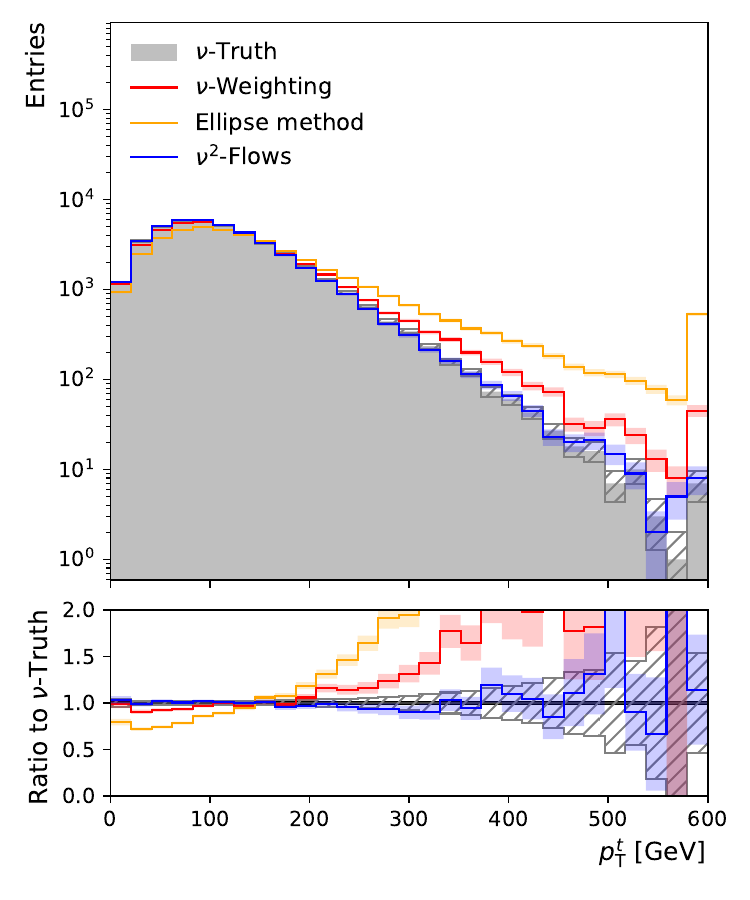}
    \caption{The reconstructed invariant mass of \Wbosons (left) and top quarks (middle), as well as the top quark \pt (right) when using the three neutrino reconstruction methods in comparison to \vtruth (shaded grey).
    }
    \label{fig:Wtmasspt}
\end{figure*}

From the reconstructed neutrinos it is also possible to reconstruct the \Wbosons, top quarks and the full \ttbar system in the event.
\Cref{fig:Wtmasspt} shows the reconstructed invariant mass of the \Wboson and top quark.
Here we always use the perfect association of jets and leptons to the two top quarks.
The strong bias to the top quark and \Wboson masses used in \cref{eq:mass_constraints} for the \vweight and \ellipse methods are clearly visible. 
In comparison, \vvflows follows the full underlying target distribution for the \Wboson mass and better captures the distribution in the reconstructed top-quark mass, but with a lower resolution than \vtruth.


\begin{figure*}[tbp]
    \centering
    \includegraphics[width=0.32\textwidth]{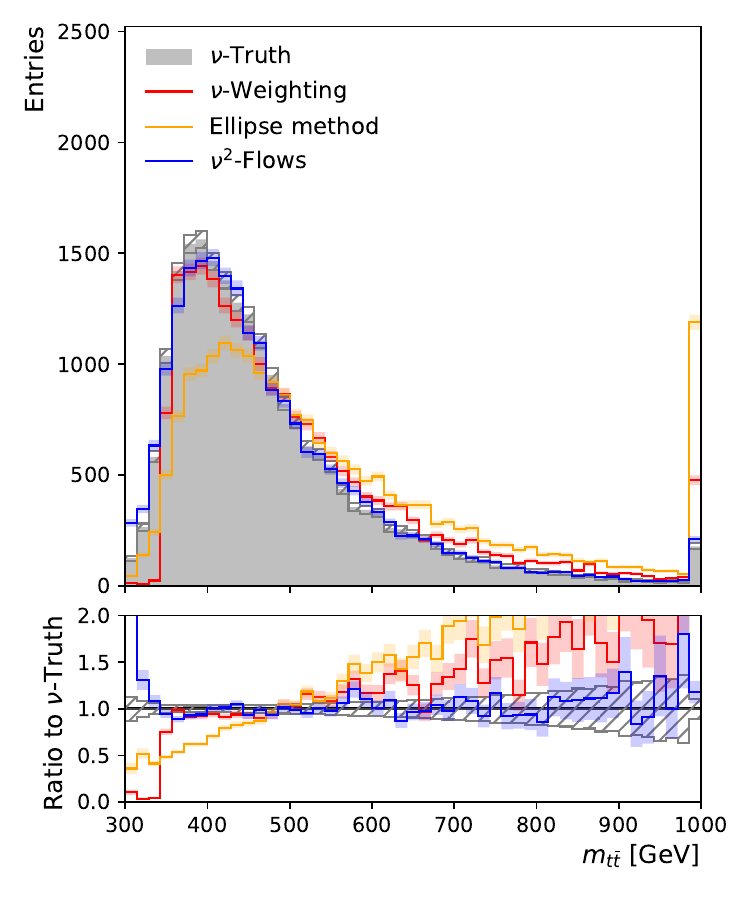}
    \includegraphics[width=0.32\textwidth]{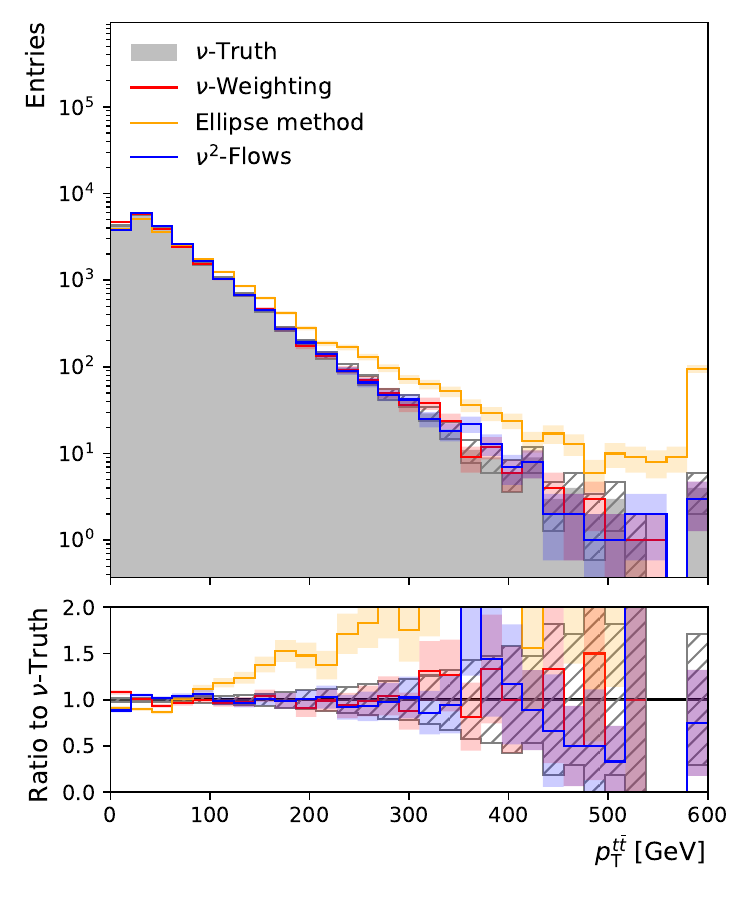}
    \includegraphics[width=0.32\textwidth]{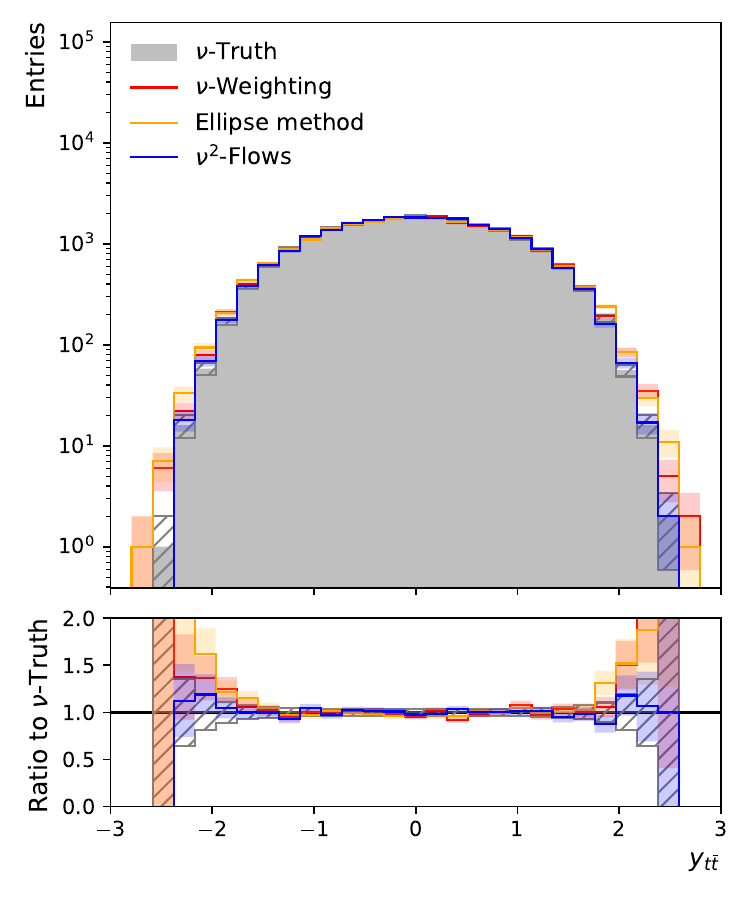}

    \caption{The invariant mass, \pt, and rapidity of the reconstructed \ttbar system when using the three neutrino reconstruction methods in comparison to \vtruth (shaded grey).
    }
    \label{fig:ttbar_kin}
\end{figure*}

However, due to the dependence on the masses of the two particles for \vweight and \ellipse it is more interesting to compare the reconstructed kinematics of the individual top quarks and the \ttbar pair.
The reconstructed top quark \pt is also compared for the three methods in \cref{fig:Wtmasspt}.
The true distribution is reproduced with \vvflows and up to approximately 200~GeV by \vweight.
However, \ellipse tends to reconstruct top quarks with a harder \pt.
In \cref{fig:ttbar_kin} the invariant mass, \pt and the rapidity of the reconstructed \ttbar system \ytt are shown.
Although discrepancies are observed at low \mttbar values, \vvflows is able to closely reproduce the kinematics of the \ttbar system much better than \vweight and \ellipse.
The \pttt and \ytt distributions are well reconstructed with \vweight but there is an overestimate in the tail of the \mttbar distribution, whereas \ellipse shows poor modelling in all three observables.


As also performed in Ref.~\cite{vflows}, we perform a cross check on the benefit of using the normalizing flow in the \vvflows architecture.
We train the \vvflows architecture but without the flow and predict the two neutrino momenta directly, establishing a simple machine learning baseline.
The performance achieved is substantially worse with strong biases in all neutrino kinematics and resulting event level distributions. 

One of the main drawbacks of \vweight, and why \ellipse is often considered despite the reduced performance, is the computational resources required.
In comparison, \vflows requires only a single forward pass for each event. The typical inference times on a CPU for single event inference are around 70~ms, with the computation time decreasing substantially with parallelised execution on a GPU as summarised in \cref{tab:inf_times}.

\begin{table}[htbp]
    \caption{Required time for single event inference using \vvflows. Times representative of using a single core of an AMD EPYC 7742 2.25GHz CPU and an \nvidia~RTX~3080 graphics card.}
    \label{tab:inf_times}
    \begin{tabular}{c c c}
        \toprule
        Resource & Batch size & Time/event [ms]\\
        \midrule
        CPU & 1 & 71.4\\
        \multirow{2}{*}{GPU} & 1 & 33.3\\
         & 1000 & 0.03\\ 
        \bottomrule
    \end{tabular}
\end{table}

  \subsection{Unfolding analysis}
  In order to evaluate the downstream impact of the improved neutrino reconstruction from \vflows, we follow the unfolding analysis performed in Ref.~\cite{ATLAS:2016pbv},
where a double differential cross section measurement is performed to measure the spin correlation in \ttbar events,
by measuring the invariant mass of the \ttbar system \mttbar and the angular separation between the two leptons \dphill.
Reconstruction of the two neutrinos is crucial in order to fully reconstruct the \ttbar system, which in Ref.~\cite{ATLAS:2016pbv} is performed using \vweight.
To benchmark our model, we replace the neutrino reconstruction with the result from \vvflows.
In addition to \vweight we also compare the performance to the \ellipse method due to its reduced computational complexity.

We focus on the reconstruction of individual observables dependent on the neutrino kinematics as well as the statistical precision of the unfolded distributions.
In addition to \dphill, we look at other observables in conjunction with \mttbar, motivated by the distributions measured in Ref.~\cite{CMS:2017iqf}.
These observables are described in \cref{tab:diff_observables} and the corresponding bin edges for the double differential unfolding and corresponding response matrices are shown in \cref{tab:diff_observables_binning}.


\begin{table}[htbp]
    \centering
    \caption{Kinematic observables of the reconstructed \ttbar system studied for an unfolding analysis in dilepton events.}
    \label{tab:diff_observables}
    \begin{tabular}{c  l}
        \toprule
        \multicolumn{2}{c}{Observables}\\
        \midrule
        \mttbar & Invariant mass of \ttbar system\\
        \dphill & Separation in $\phi$ between the two leptons\\
        \pttop & Transverse momentum of the top quark\\
        \pttt & Transverse momentum of the \ttbar system\\
        \ytt & Rapidity of the \ttbar system\\
        \bottomrule
    \end{tabular}
\end{table}

\begin{table}[htbp]
    \centering
    \caption{Bin edges used for each of the kinematic observables of the reconstructed \ttbar system studied for two dimensional unfolding analyses.}
    \label{tab:diff_observables_binning}
    \begin{tabular}{c r l}
        \toprule
        Observable & Bin edges\\
        \midrule
        \mttbar & [0, 400, 500, 800, inf] & GeV\\
        \dphill & [0.0, 0.25, 0.5, 0.75, 1.0] & rad/$\pi$\\
        \pttop  & [0, 75, 125, 175, inf] & GeV\\
        \pttt   & [0, 70, 140, 200, inf] & GeV\\
        \ytt    & [-inf, -1.0, 0.0, 1.0, inf] &\\
        \bottomrule
    \end{tabular}
\end{table}

All distributions are compared to \vtruth, and the total uncertainty in each bin after unfolding is calculated with respect to the optimal performance achieved when using \vtruth.
The correct jet and lepton association is used for both top quarks in order to remove the effects arising from matching inefficiencies.
We perform the unfolding using the Singular Value Decomposition (SVD) method~\cite{Hocker:1995kb} with a regularisation factor of 7, using the implementation provided in \texttt{RooUnfold}~\cite{Brenner:2019lmf}.
The regularisation factor was optimised for the \vtruth distributions for a reduced $\chi^2$ value closest in agreement to one for the four double differential distributions.
In all cases we only consider the \ttbar process and ignore the impact of background estimation and subtraction in the methods.

Although more modern machine learning approaches for unfolding are an active area of research~\cite{Gagunashvili:2010zw,Datta:2018mwd,Andreassen:2019cjw,Bellagente:2020piv,Backes:2022vmn,Chan:2023tbf}, we leave their study and application to dilepton \ttbar events to future studies, in particular the study of unbinned multidimensional unfolding with the reconstructed neutrino kinematics.

\begin{figure*}[htp]
    \includegraphics[width=0.32\textwidth]{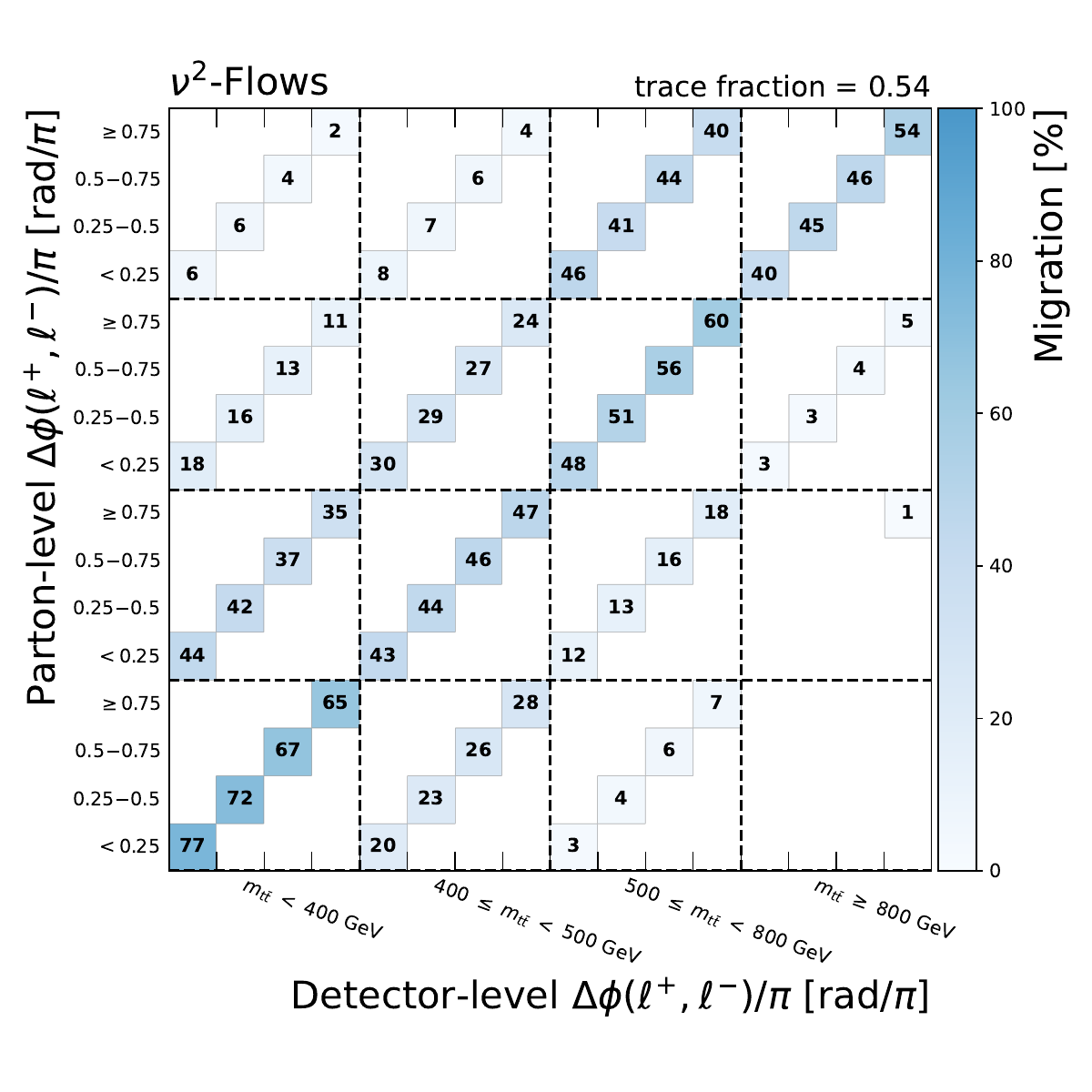}
    \includegraphics[width=0.32\textwidth]{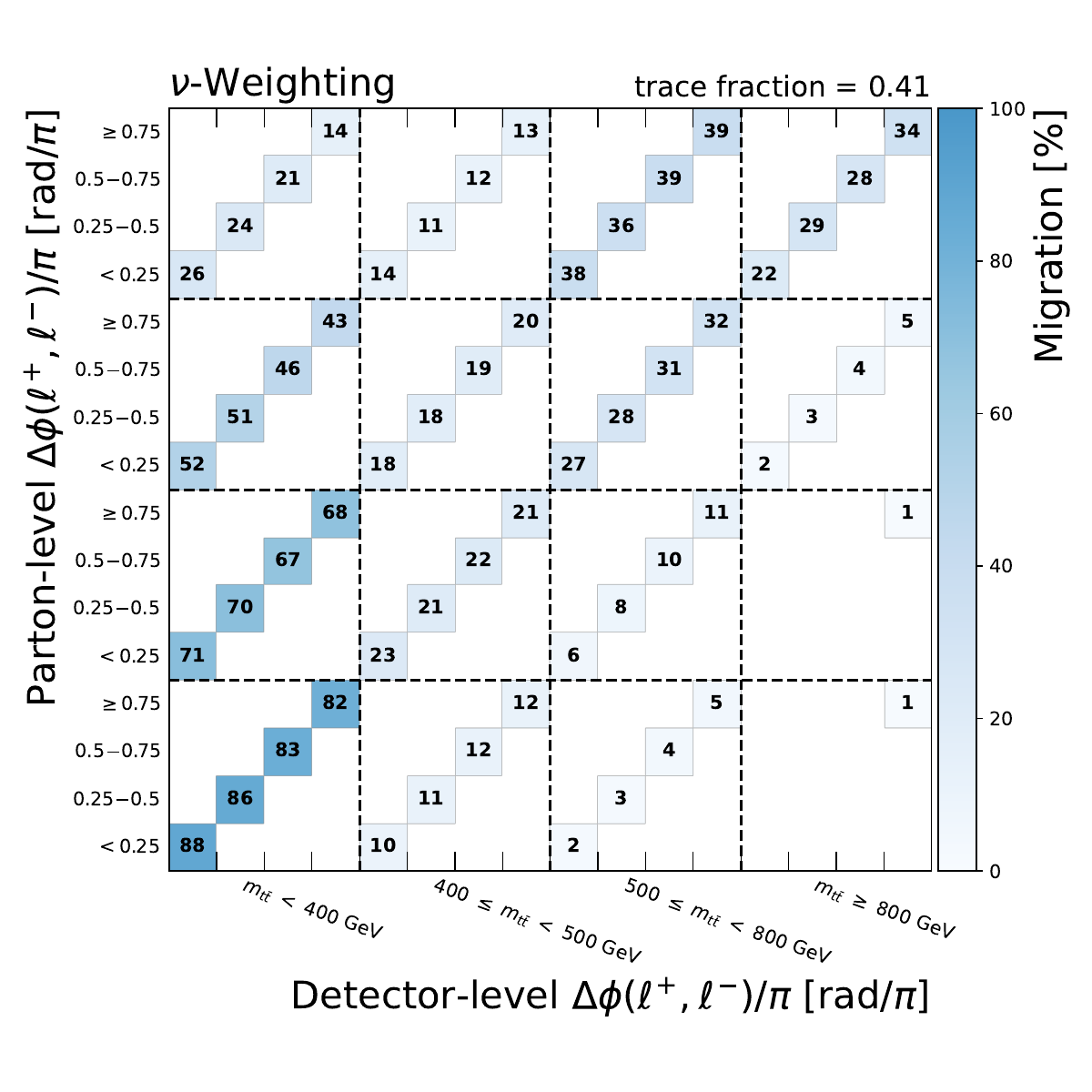}
    \includegraphics[width=0.32\textwidth]{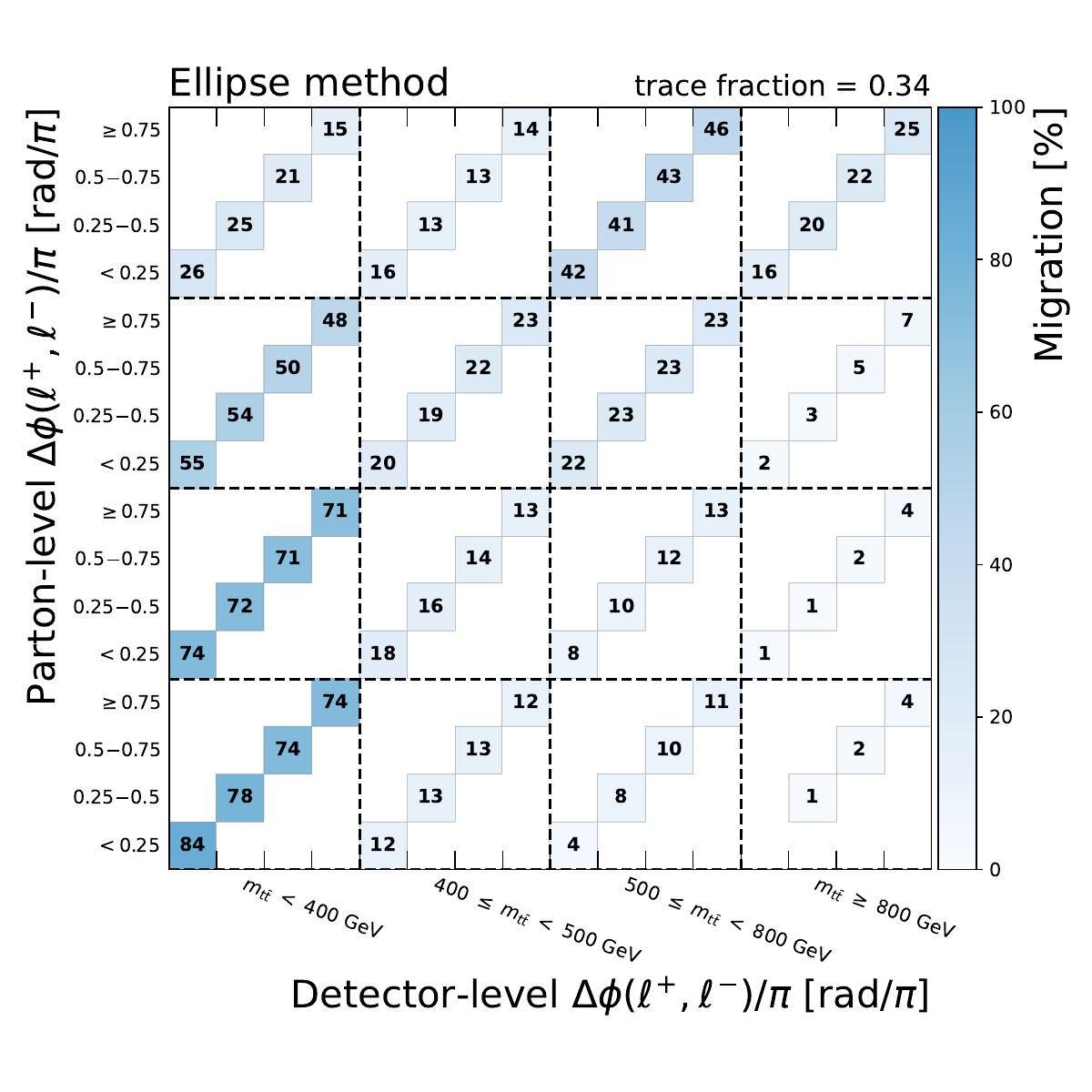}
    \caption{Binned response matrices for the double differential measurement of \mttbar and \dphill when using each of the three methods for neutrino reconstruction. The binning is symmetric for both the parton and detector level observables, however the \mttbar bins are labelled on the $x$-axis with the \dphill bins labelled on the $y$-axis.
    The trace fraction is calculated for each method for a simple quantitative comparison and is 0.73 when using \vtruth.}
    \label{fig:unfold_dphill}
\end{figure*}

\begin{figure*}[htp]
    \includegraphics[width=0.32\textwidth]{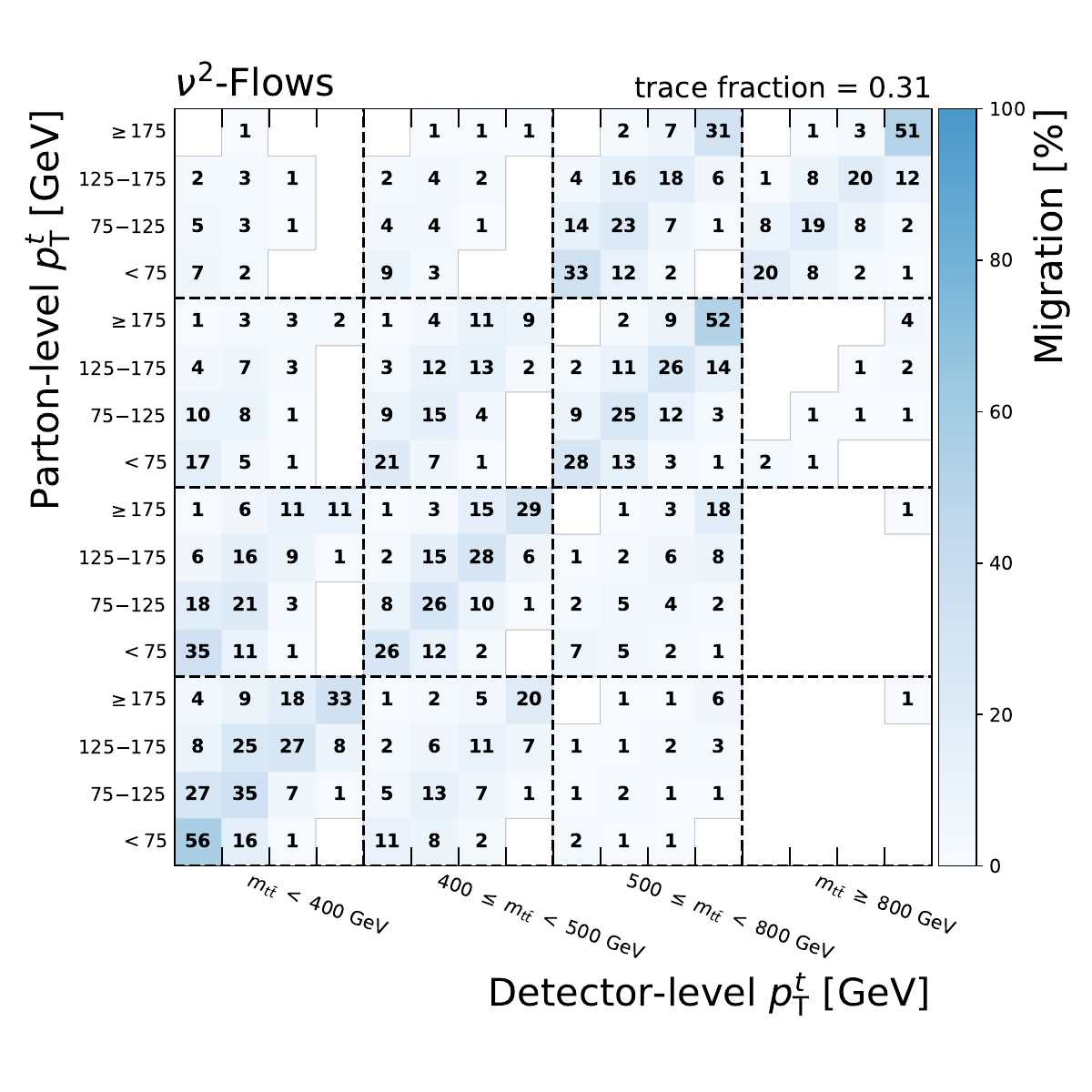}
    \includegraphics[width=0.32\textwidth]{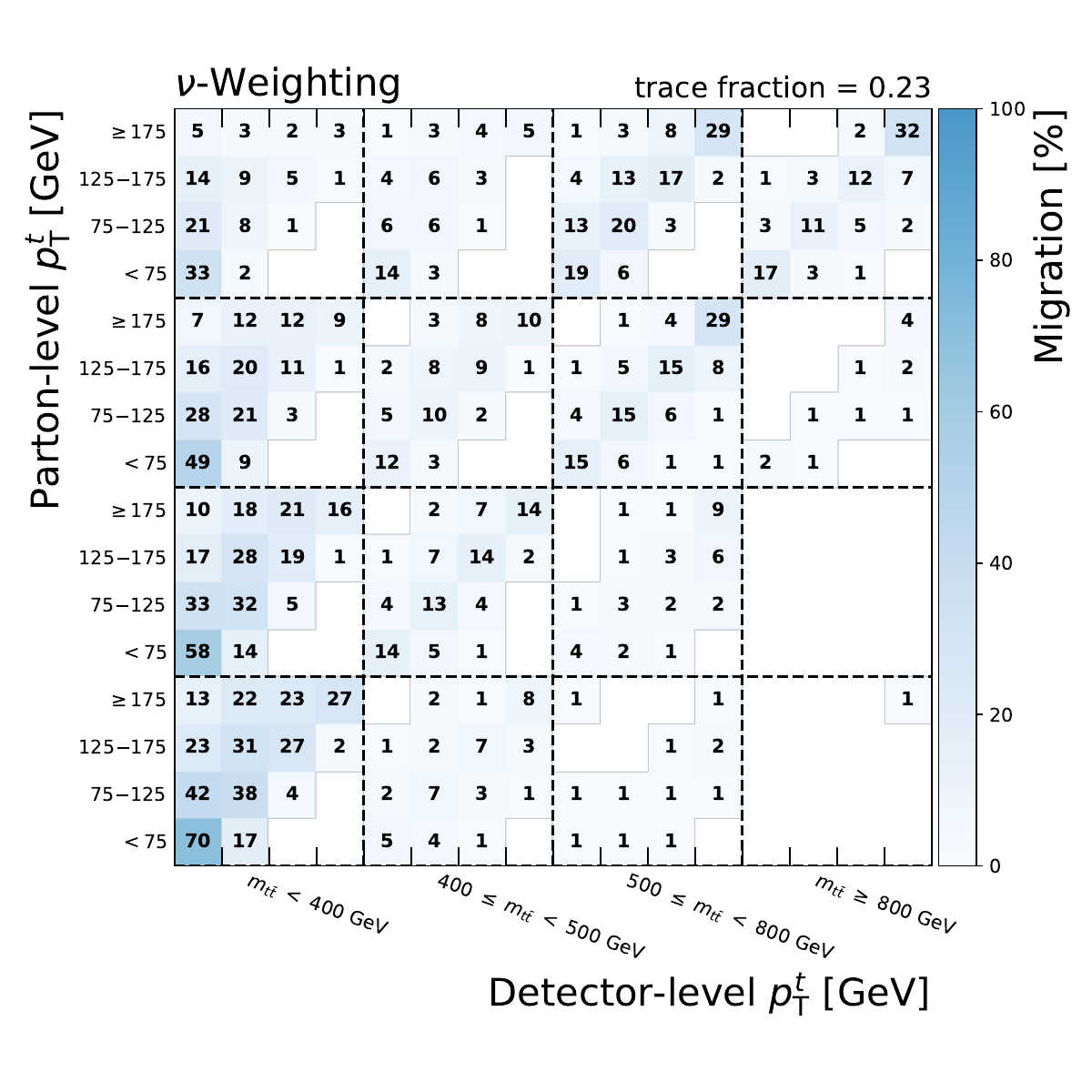}
    \includegraphics[width=0.32\textwidth]{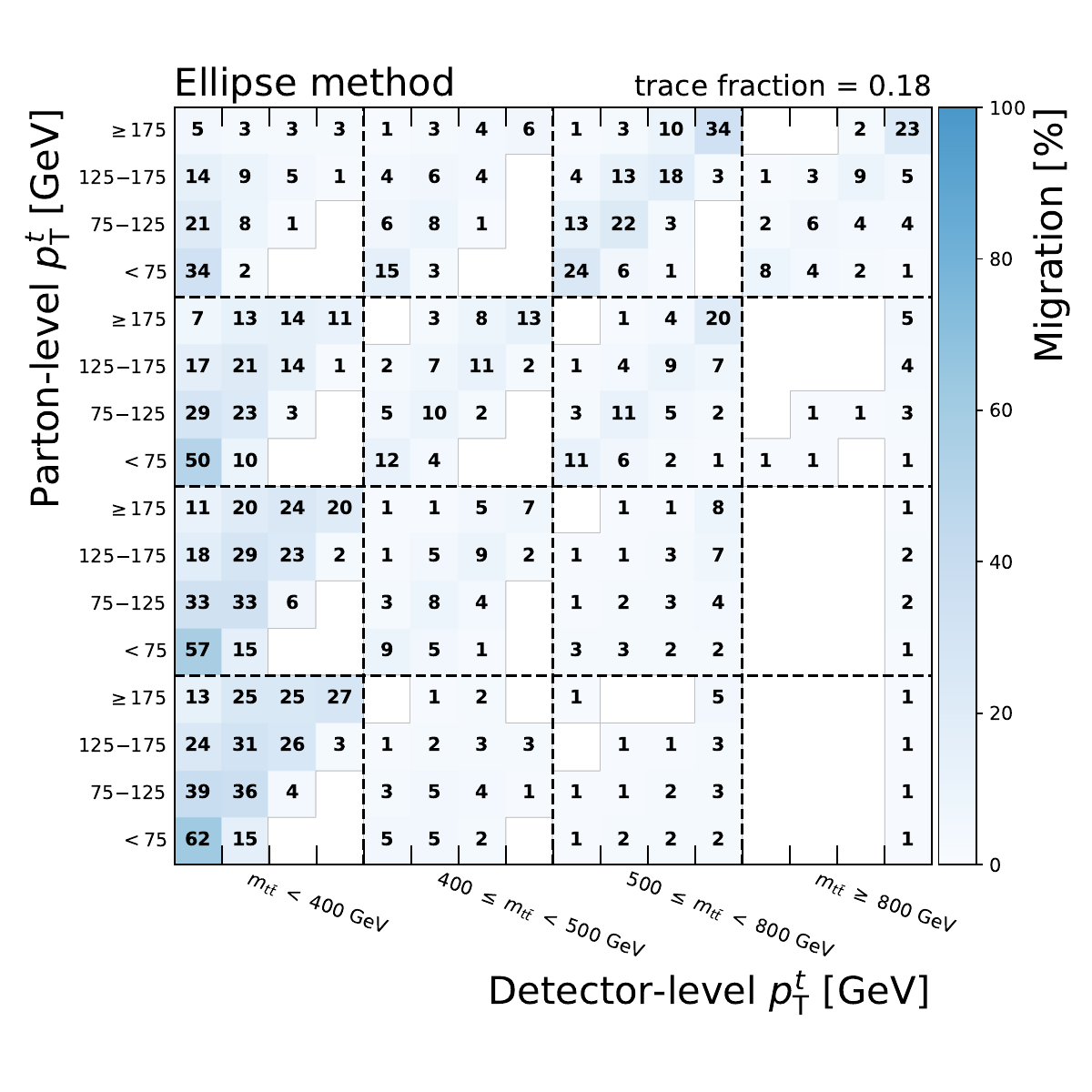}
    \caption{Binned response matrices for the double differential measurement of \mttbar and \pttop when using each of the three methods for neutrino reconstruction. The binning is symmetric for both the parton and detector level observables, however the \mttbar bins are labelled on the $x$-axis with the \pttop bins labelled on the $y$-axis.
    The trace fraction is calculated for each method for a simple quantitative comparison and is 0.62 when using \vtruth.}
    \label{fig:unfold_pttop}
\end{figure*}

The response matrix using each of the neutrino reconstruction methods is shown in \cref{fig:unfold_dphill} for the two dimensional binning in \mttbar and \dphill and in \cref{fig:unfold_pttop} for \mttbar and \pttop.
In the ideal case only the main diagonal would contain entries, however due to inefficiencies in the neutrino reconstruction methods as well as detector resolution effects, off diagonal elements are unavoidable.
In both cases it is clear that using \vvflows to reconstruct the neutrino pair results in a more diagonal response matrix than the other two approaches.
This is quantified by the trace fraction of each matrix.

Although the trace fraction can give a good measure of which method is performing best, the off diagonal elements still contribute to the unfolded distributions.
To quantitatively assess the true impact of using each method, the response matrices are inverted using SVD and the overall uncertainties for each bin at parton level are calculated.

\begin{figure*}[htpb]
    \centering
    \includegraphics[width=0.48\textwidth]{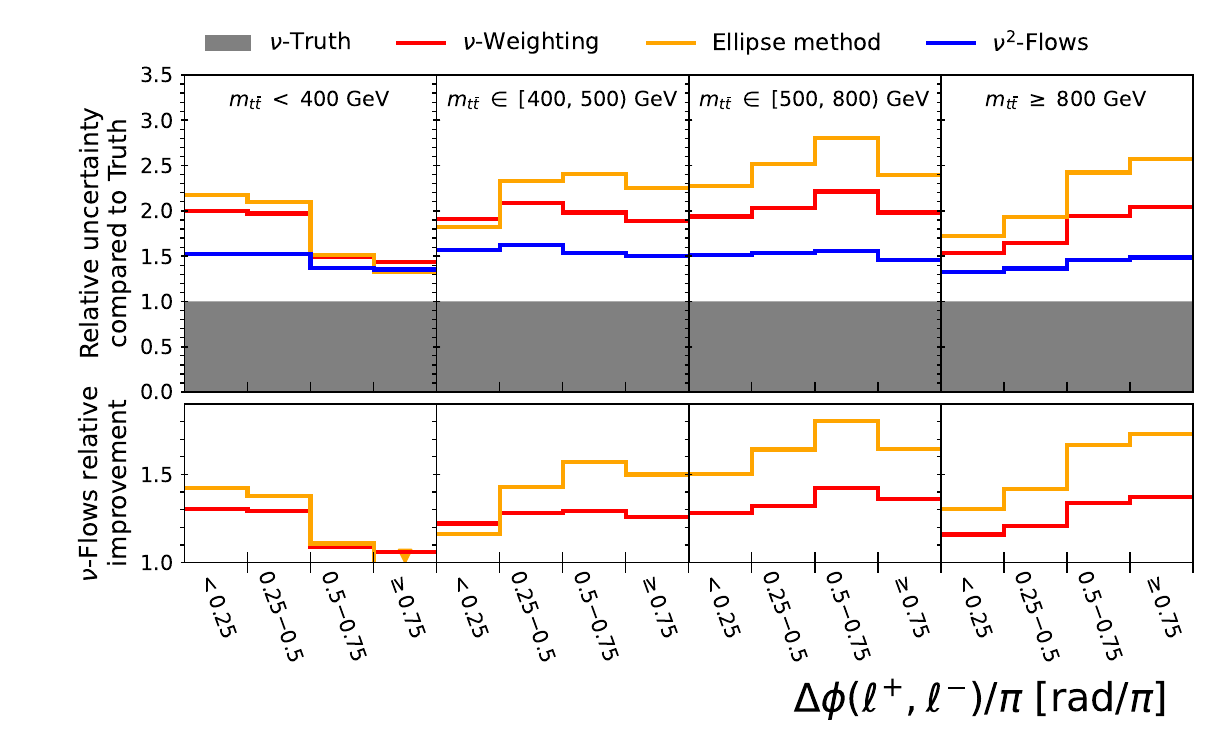}
    \includegraphics[width=0.48\textwidth]{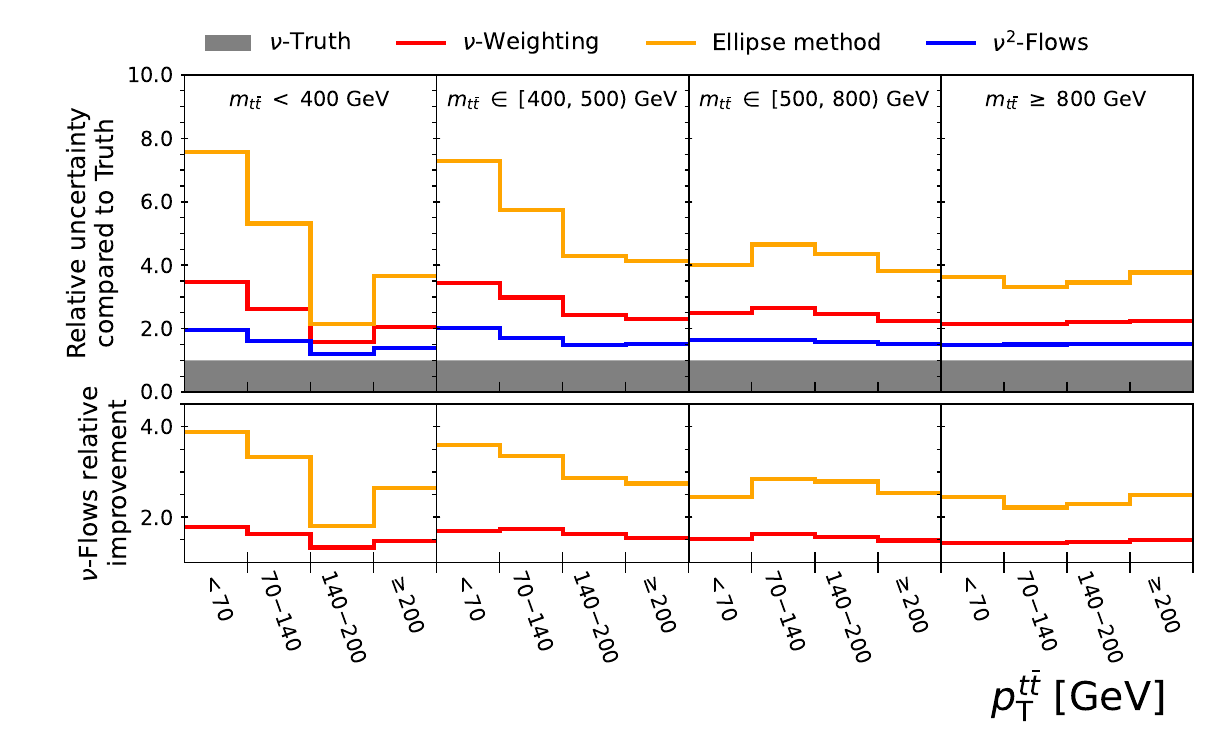} \\
    \includegraphics[width=0.48\textwidth]{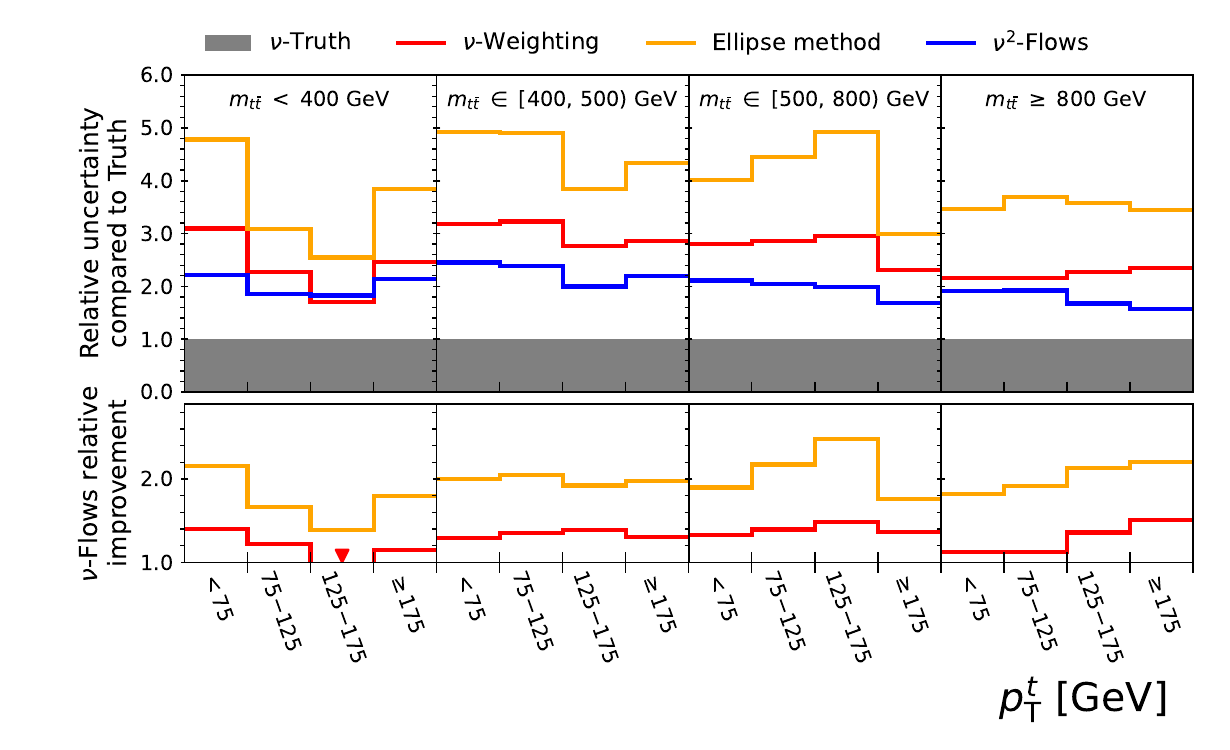}
    \includegraphics[width=0.48\textwidth]{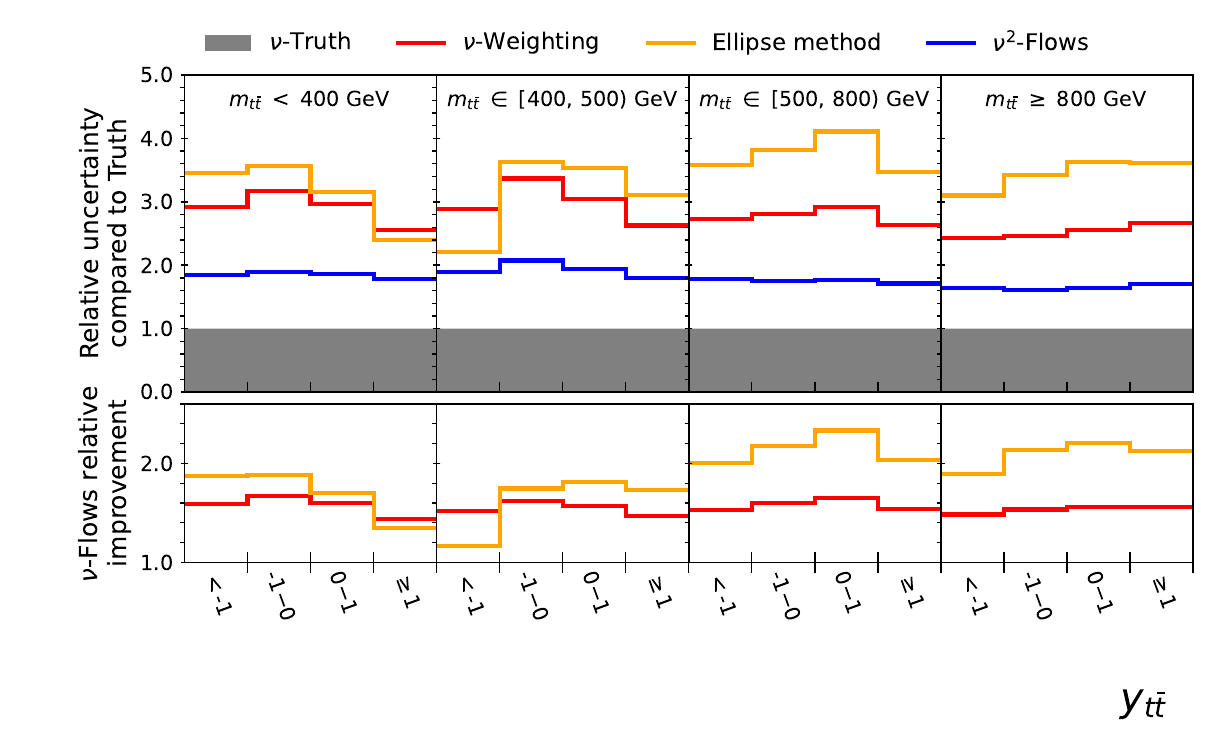}
    \caption{The relative statistical uncertainty in each unfolded bin of the two dimensional distributions for \mttbar and \dphill (top left), \pttop (top right), \pttt (bottom left), and \ytt (bottom right).
    Comparing the three reconstruction methods with respect to \vtruth (upper pad), and the relative improvement of \vflows with respect to \vweight and \ellipse (lower pad).}
    \label{fig:stat_gain_unfold_all}
\end{figure*}

\Cref{fig:stat_gain_unfold_all} shows the relative statistical uncertainty for each method with respect to \vtruth.
The values are obtained for each bin of the unfolded distributions using SVD with the chosen level of regularisation.
For the individual bins of the four double differential distributions the uncertainties are typically a factor of 1.5 to two times smaller when using \vvflows compared to \vweight, and up to four times smaller in comparison with \ellipse.



  \subsection{Robustness to training sample}
  In comparison to the standard analytical approaches, \vflows is trained on a specific sample of Monte Carlo simulated events.
This could introduce a performance dependence on the sample used for training, which may not be optimal for all generators.
It should be noted that the same model is used for all events and, just like the analytical approaches, is independent of which samples it is applied to.
However, if \vflows has learned sample specific effects this can lead to a suboptimal performance or even unusable levels of performance when applied to other samples.

To study the impact of this effect we train \vvflows using the alternative \ttbar dilepton sample (\vvflowsPy) and use it to reconstruct the neutrinos for the nominal \ttbar sample.
We compare the reconstructed kinematic distributions as well as the uncertainties in each bin of the unfolded distributions.


Negligible differences are observed in the reconstructed neutrino kinematics, though the difference can clearly be seen for the reconstructed \Wboson mass and a slight difference is also seen for the reconstructed top quark invariant mass.

\begin{figure*}[htbp]
\centering
\includegraphics[width=0.32\textwidth]{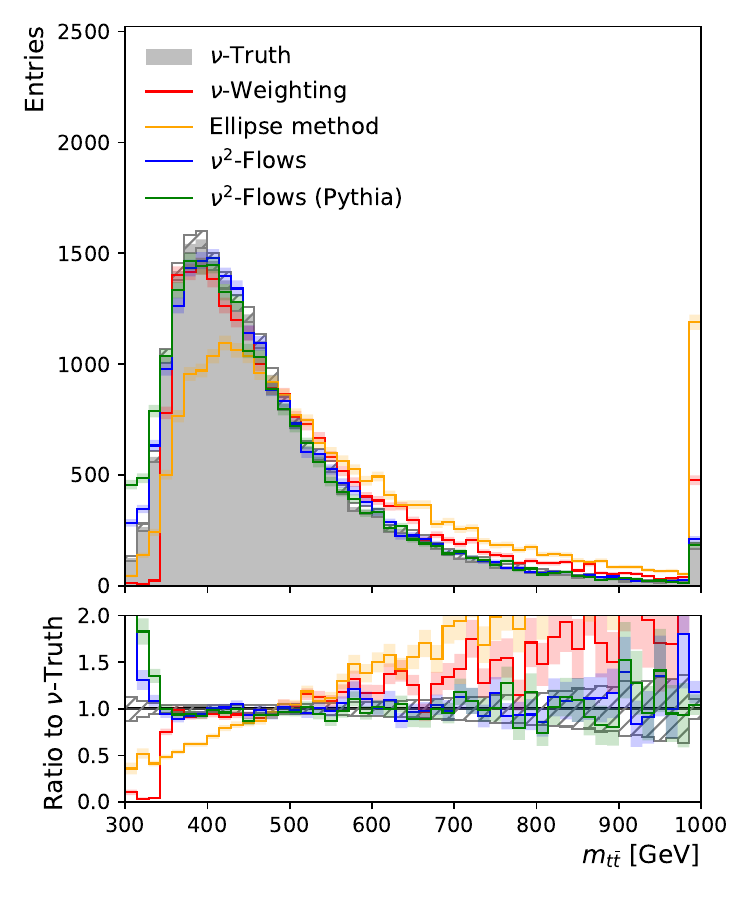}
\includegraphics[width=0.32\textwidth]{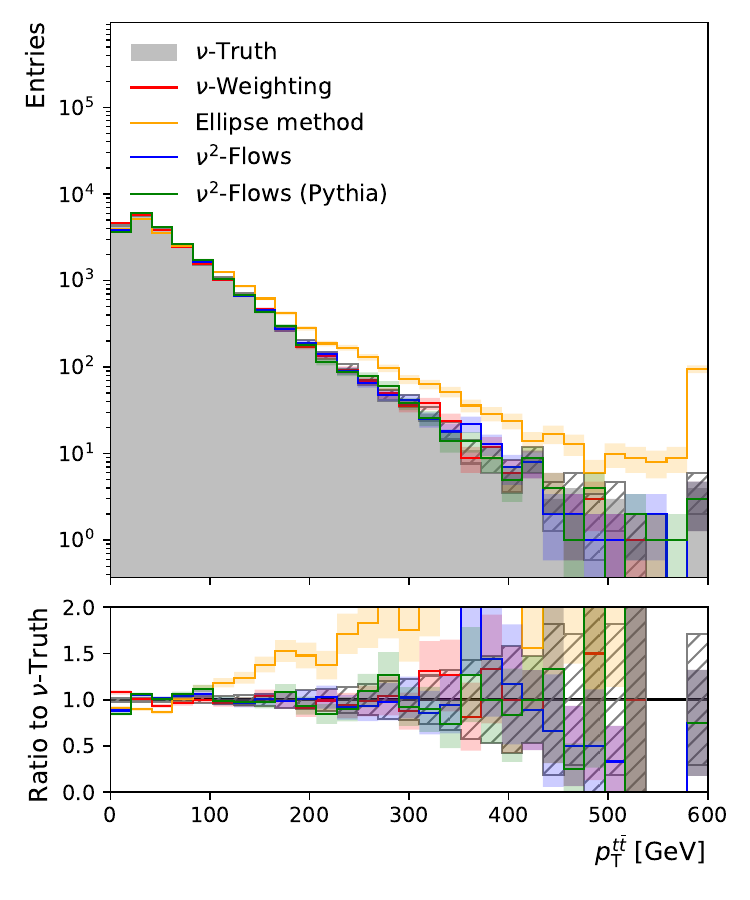}
\includegraphics[width=0.32\textwidth]{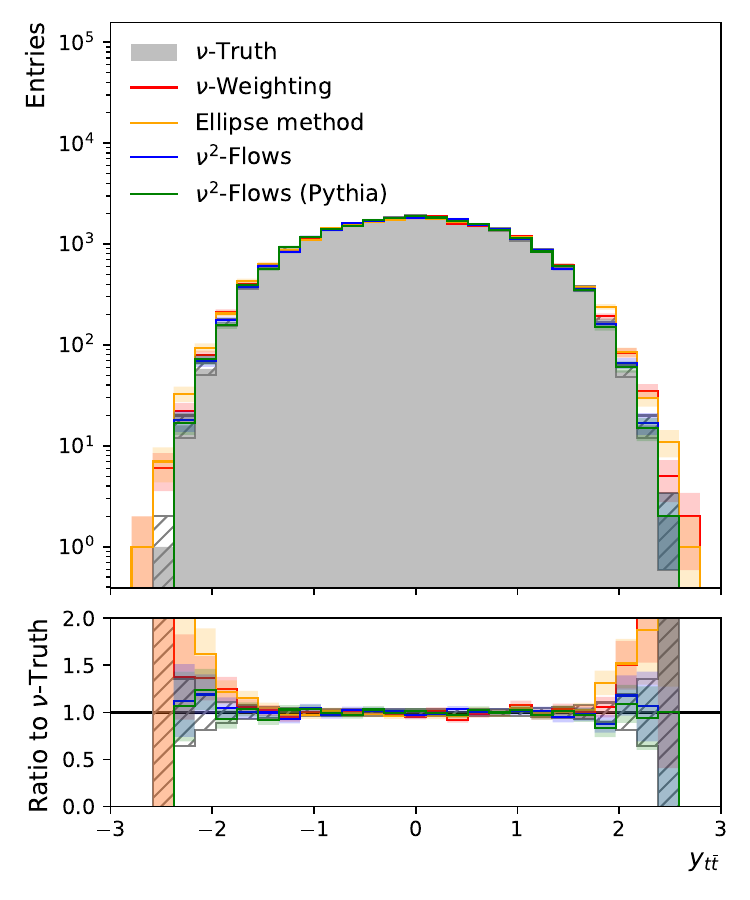}

\caption{The invariant mass, \pt, and rapidity of the reconstructed \ttbar system when using \vvflows trained on the nominal or alternative sample, in comparison to the two baseline approaches and \vtruth (shaded grey).
}
\label{fig:pythiattbar}
\end{figure*}

\begin{figure}[htbp]
        \includegraphics[width=0.47\textwidth]{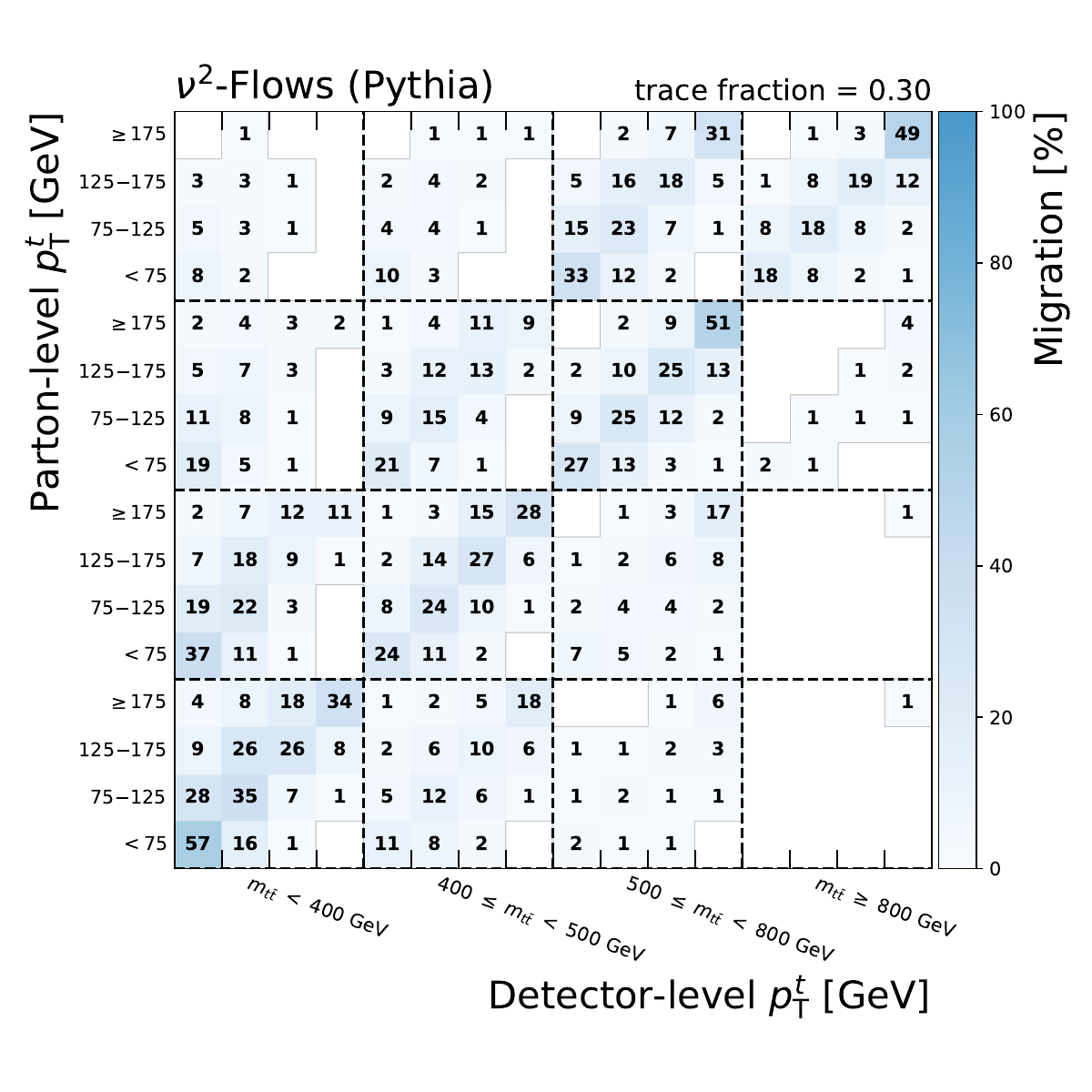}
        \caption{Binned response matrices for the double differential measurement of \mttbar and \pttop when using \vvflowsPy but evaluating on the nominal \ttbar sample. The binning is symmetric for both the parton and detector level observables, however the \mttbar bins are labelled on the $x$-axis with the \pttop bins labelled on the $y$-axis.
        The trace fraction is calculated for each method for a simple quantitative comparison and is 0.62 when using \vtruth and 0.31 for the nominal \vvflows.}
        \label{fig:unfold_pythia}
\end{figure}

Some small differences are also observed in the tails of the reconstructed top quark and \ttbar properties in \cref{fig:pythiattbar}, however the performance is still substantially improved in comparison to \vweight and \ellipse.
The response matrix for \vvflowsPy for the double differential distribution in \mttbar and \pttop is shown in \cref{fig:unfold_pythia}.
These differences translate to a very slight change in the statistical precision in each bin after performing the unfolding.

As another test of robustness we study how well \vvflows performs in comparison the \vweight and \ellipse at reconstructing \ttbar events which have been simulated with additional initial state radiation~(ISR).
This sample has 280,000 events and is otherwise the same as the nominal sample, however substantial differences are be observed in the jet multiplicity distributions and underlying kinematics of the top-quark and \ttbar pair.
Most notably the distribution of \pttt is much harder, due to the recoil against the additional ISR.

\begin{figure*}[htbp]
        \centering
        \includegraphics[width=0.32\textwidth]{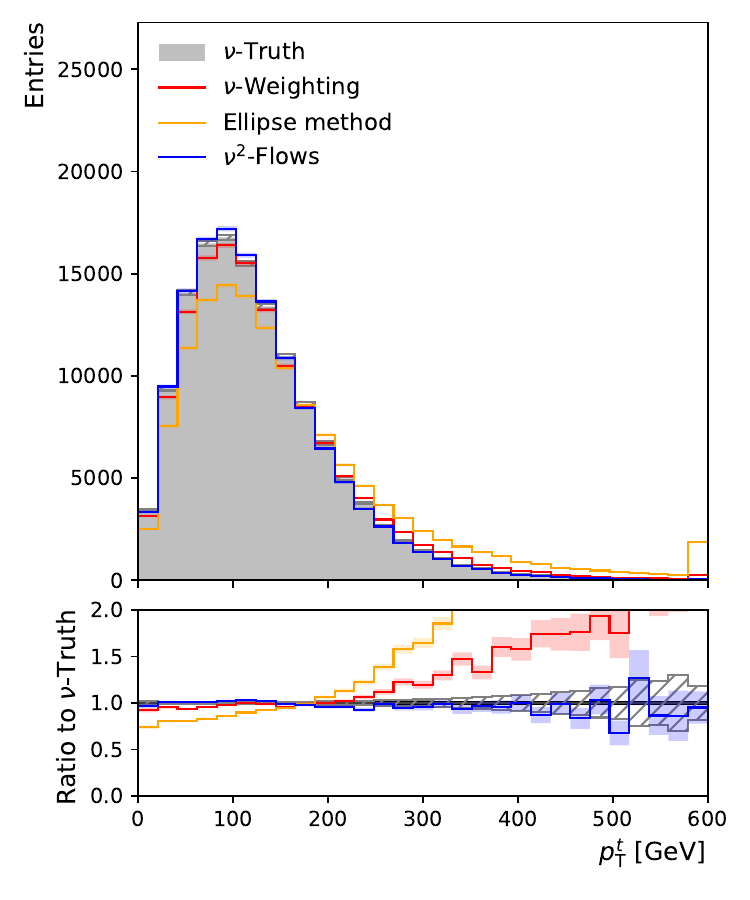}
        \includegraphics[width=0.32\textwidth]{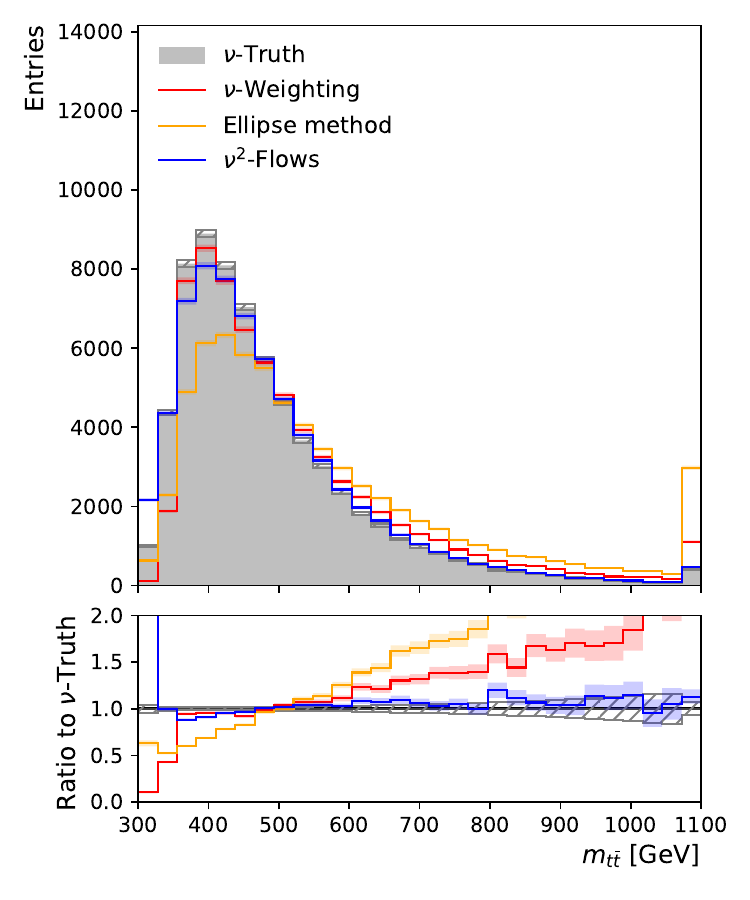}
        \includegraphics[width=0.32\textwidth]{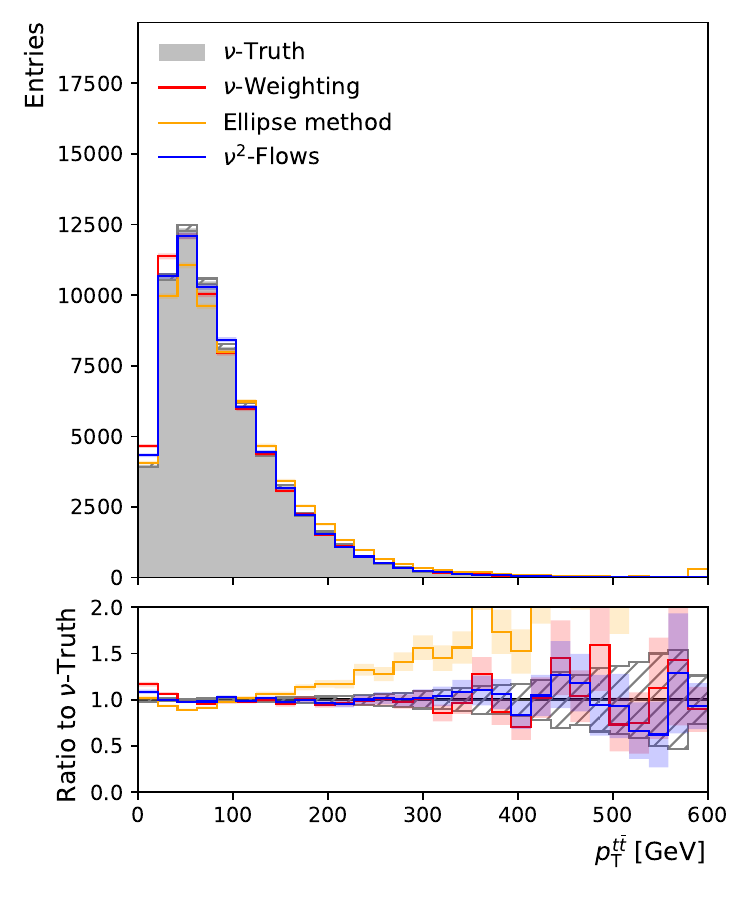}
        
        \caption{The reconstructed top quark \pt, and the invariant mass and \pt of the reconstructed \ttbar system for an independent sample with additional initial state radiation when using \vvflows trained on the nominal sample in comparison to the two baseline approaches and \vtruth (shaded grey).
        }
        \label{fig:extraISRttbar}
\end{figure*}

\Cref{fig:extraISRttbar} shows the reconstructed distributions for the three approaches.
As before, \vvflows exhibits very good agreement across the majority of the kinematic phase space, despite having been optimised for a different sample.

  \subsection{Measuring the top quark mass}
  In contrast to \vweight and \ellipse, which both assume values for the top quark mass in the neutrino reconstruction,
\vvflows only implicitly learns this relation from the training data. 
To test whether \vvflows has sensitivity to the underlying top quark mass, we use the default \vvflows model to evaluate events in which the truth $m_t$ has been changed to either 171~GeV or 175~GeV.
These samples each have 160,000 events and are otherwise the same as the nominal sample.

We compare the reconstructed top quark mass using the four momenta of the lepton, $b$-quark and neutrino ($m_t$) to the invariant mass using only of the lepton and $b$-quark from the same top decay ($m_{b\ell}$).

\begin{figure*}[hbtp]
    \includegraphics[width=0.32\textwidth]{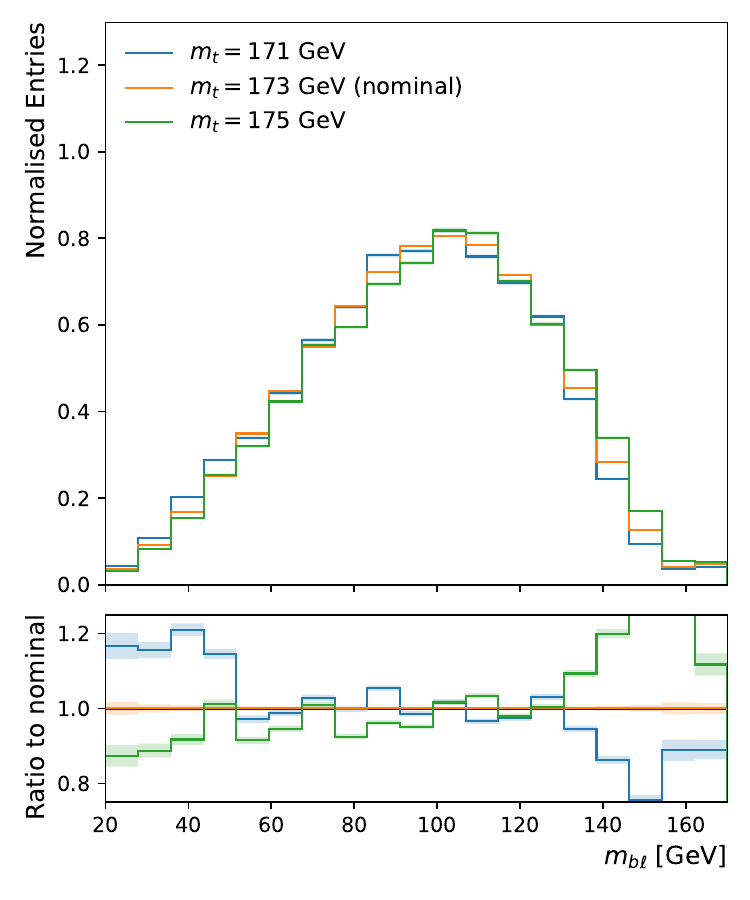}
    \includegraphics[width=0.32\textwidth]{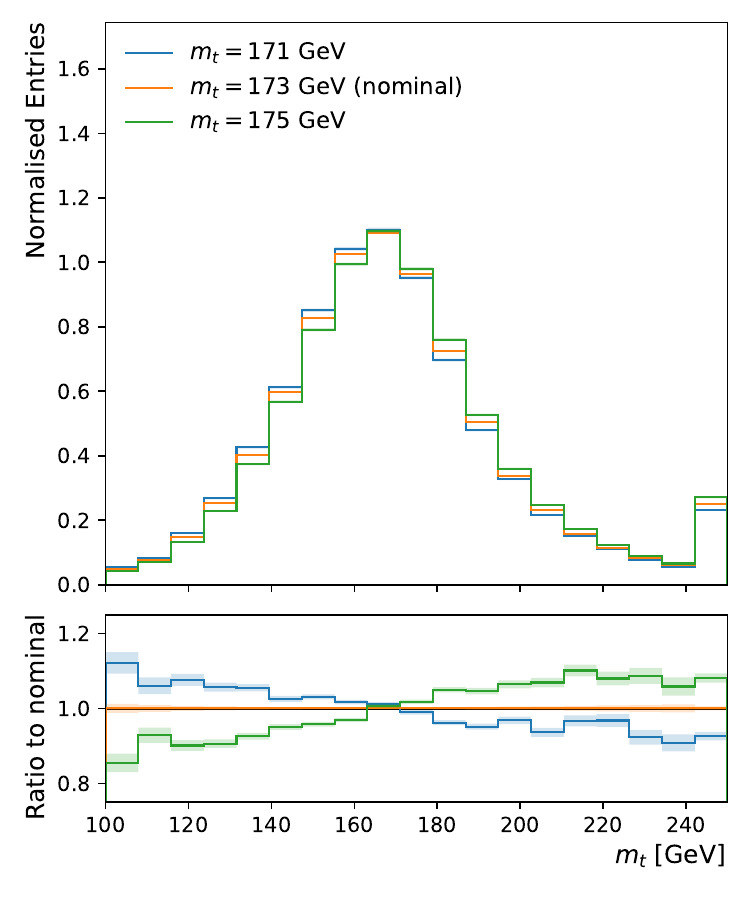}
    \includegraphics[width=0.32\textwidth]{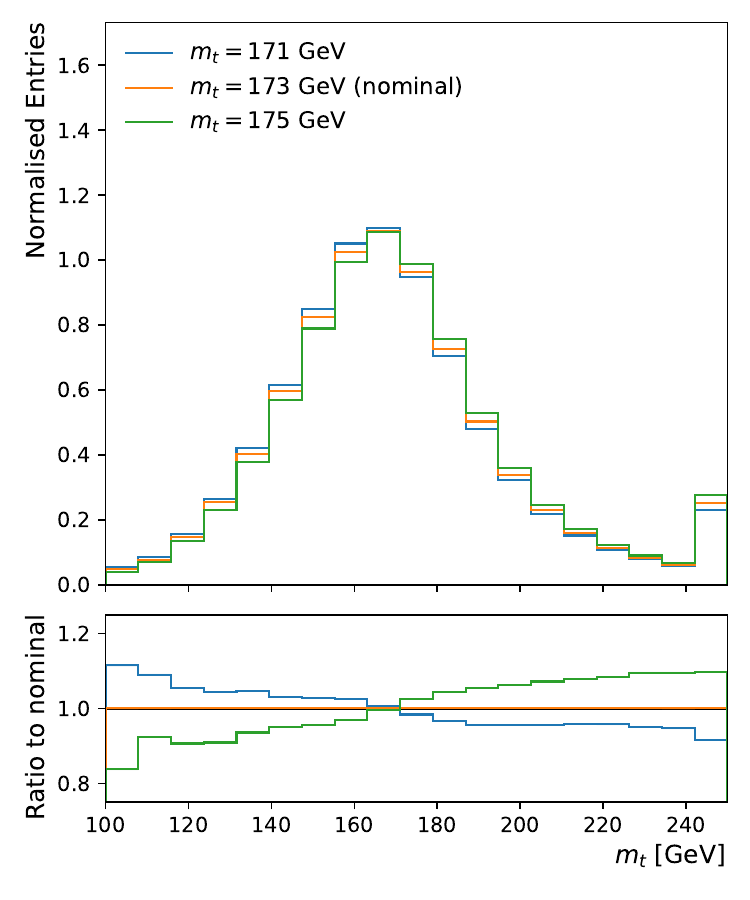}
    \caption{The reconstructed top quark mass using just from the lepton and $b$-quark from the top quark decay ($m_{b\ell}$, left), the full invariant mass using \vvflows to reconstruct the neutrinos from the top quarks ($m_t$, middle), and a smoothed template for $m_t$ obtained by sampling 256 solutions for each event with \vvflows (right).
    The statistical uncertainty on the distributions arising from the training dataset has not been propagated into the final statistical uncertainty.}
    \label{fig:top_mass_reco}
\end{figure*}

The distributions are shown in \cref{fig:top_mass_reco}. 
Despite training purely on events with the nominal top quark mass, the reconstructed distribution using \vvflows is sensitive to the difference in truth $m_t$. 
The separation between the three templates is similar to using $m_{b\ell}$, however for \vvflows the difference is more prominent in the bulk of the distribution. 
This sensitivity could be improved by training \vvflows on samples with a range of values for $m_t$ and parametrising the network, however this would introduce additional computational complexity to the method.
Another benefit of \vvflows is also demonstrated, with a smoother templates constructed by sampling multiple neutrino solutions for each event.

  \section{Conclusions}
  With \vvflows we have built upon the success of the \vflows method to employ conditional normalizing flows to reconstruct the momentum vectors of multiple neutrinos in a single event.
In comparison to other commonly used approaches, \vvflows is able to reconstruct both neutrinos without enforcing strong constraints on reconstructed particle masses in order to find solutions in an under-constrained system.
This translates to a reduced bias in the reconstruction of neutrinos, without a preference for back-to-back neutrinos, and with a more accurate reconstruction of the kinematics of individual top quarks and the full \ttbar system.
The reconstructed neutrinos can be used directly or potentially combined with other machine learning approaches which aim to reconstruct the underlying hard scatter event~\cite{Spatter,SPANet,Badea:2022dzb,Qiu:2023ihi,Ehrke:2023cpn}.
The generalised architecture introduced in \vvflows has been designed to be easy to extend to any neutrino multiplicity, and does not place restrictions on the multiplicities of reconstructed objects, or how they are combined to extract information from the event.

In applying \vvflows to dilepton \ttbar events 
we achieve significant improvements in the statistical precision of unfolded distributions in comparison to the most commonly used analytical approaches for a wide range of distributions of interest.
These improvements can lead to an increased sensitivity in the measurement of top quark spin correlation and entanglement at the LHC.
Furthermore, in contrast to other approaches, \vvflows has solutions for all events and the fast single event inference below 75~ms on a single computing core.
As each sample is associated with a probability from the transformation under the normalizing flow, this could also provide opportunities for separating \ttbar events from background processes, similar to the weight in \vweight.
This could be done using the probability of the single solution per event, or, as multiple solutions can be sampled for each event, the highest possible probability value for an event.


  \section*{Acknowledgements}
  The authors would like to thank Chris Pollard for support producing the Pythia8 standalone samples.

The authors acknowledge funding through the SNSF Sinergia grant ``Robust Deep Density Models for High-Energy Particle Physics and Solar Flare Analysis (RODEM)'' with funding number CRSII$5\_193716$, and the SNSF project grant 200020\_212127 ``At the two upgrade frontiers: machine learning and the ITk Pixel detector''.
They would also like to acknowledge individual funding acquired through the Swiss Government Excellence Scholarships for Foreign Scholars and the Feodor Lynen Research Fellowship from the Alexander von Humboldt foundation.

JAR and ML contributed equally to this work.

  \bibliography{main.bib}

  \appendix

  \section{Additional material}
  
  \subsection{Unfolding efficiencies}

The relative uncertainty for each bin in the unfolded distributions is compared in \cref{tab:app:rel_unfold_uncerts} for all methods with respect to \vtruth, including both \vvflows and \vvflowsPy.
The efficiency of finding a solution with the \vweight and \ellipse methods for each bin of the double differential distributions are summarised in \cref{tab:app:sol_efficiency}.
As \vvflows is able to find a solution for all events, a value lower than 100\% indicates where additional statistical sensitivity can be achieved with \vvflows.
This is particularly notable with increasing values of \mttbar, but also for higher values of \pttt.
This is especially the case in comparison to \ellipse, with some bins having more than 50\% of events without solutions,
however, in certain unfolding bins 10-20\% of events also do not have solutions when using \vweight.

\begin{table*}[thp]
    \centering
    \caption{Relative uncertainty in each bin of the respective unfolded double differential distributions for each neutrino reconstruction method with respect to the uncertainty when using \vtruth.
    The bins are ordered first by increasing \mttbar followed by the second variable, with vertical dividers indicating the bin edges in \mttbar.
    The method with the smallest relative increase in uncertainty in comparison to \vtruth is highlighted in bold.}
    \label{tab:app:rel_unfold_uncerts}
    \begin{tabular}{c c r r r r r | r r r r | r r r r | r r r r}
    \toprule
    \multicolumn{2}{c}{Observables} & Method & \multicolumn{16}{c}{Relative uncertainty to \vtruth (per bin)}\\
    \midrule

    \multirow{16}{*}{\wrapentry{0.07\textwidth}{\mttbar}} & \multirow{4}{*}{\ytt}
      & \vweight & \,2.9 & \,3.2 & \,3.0 & \,2.6 & \,2.9 & \,3.4 & \,3.1 & \,2.6 & \,2.7 & \,2.8 & \,2.9 & \,2.6 & \,2.4 & \,2.5 & \,2.6 & \,2.7\\
    & & \ellipse & \,3.5 & \,3.6 & \,3.2 & \,2.4 & \,2.2 & \,3.6 & \,3.5 & \,3.1 & \,3.6 & \,3.8 & \,4.1 & \,3.5 & \,3.1 & \,3.4 & \,3.6 & \,3.6\\
    & & \vvflows & \textbf{\,1.8} & \textbf{\,1.9} & \textbf{\,1.9} & \textbf{\,1.8} & \textbf{\,1.9} & \textbf{\,2.1} & \textbf{\,1.9} & \textbf{\,1.8} & \textbf{\,1.8} & \textbf{\,1.8} & \textbf{\,1.8} & \textbf{\,1.7} & \textbf{\,1.6} & \textbf{\,1.6} & \textbf{\,1.6} & \textbf{\,1.7}\\
    & & \vvflowspy & \,1.9 & \textbf{\,1.9} & \textbf{\,1.9} & \textbf{\,1.8} & \textbf{\,1.9} & \textbf{\,2.1} & \,2.0 & \textbf{\,1.8} & \textbf{\,1.8} & \textbf{\,1.8} & \textbf{\,1.8} & \textbf{\,1.7} & \,1.7 & \textbf{\,1.6} & \,1.7 & \,1.8\\

    \cmidrule{2-19}
      & \multirow{4}{*}{\pttt}
      & \vweight & \,3.5 & \,2.6 & \,1.6 & \,2.1 & \,3.4 & \,3.0 & \,2.4 & \,2.3 & \,2.5 & \,2.7 & \,2.5 & \,2.2 & \,2.1 & \,2.2 & \,2.2 & \,2.3\\
    & & \ellipse & \,7.6 & \,5.3 & \,2.2 & \,3.7 & \,7.3 & \,5.7 & \,4.3 & \,4.1 & \,4.0 & \,4.7 & \,4.4 & \,3.8 & \,3.6 & \,3.3 & \,3.5 & \,3.8\\
    & & \vvflows & \textbf{\,1.9} & \textbf{\,1.6} & \textbf{\,1.2} & \textbf{\,1.4} & \textbf{\,2.0} & \textbf{\,1.7} & \textbf{\,1.5} & \textbf{\,1.5} & \textbf{\,1.6} & \textbf{\,1.6} & \textbf{\,1.6} & \textbf{\,1.5} & \textbf{\,1.5} & \textbf{\,1.5} & \textbf{\,1.5} & \textbf{\,1.5}\\
    & & \vvflowspy & \,2.0 & \textbf{\,1.6} & \textbf{\,1.2} & \textbf{\,1.4} & \textbf{\,2.0} & \textbf{\,1.7} & \textbf{\,1.5} & \textbf{\,1.5} & \,1.7 & \textbf{\,1.6} & \textbf{\,1.6} & \textbf{\,1.5} & \textbf{\,1.5} & \textbf{\,1.5} & \textbf{\,1.5} & \textbf{\,1.5}\\

    \cmidrule{2-19}
      & \multirow{4}{*}{\pttop}
      & \vweight & \,3.1 & \,2.3 & \textbf{\,1.7} & \,2.5 & \,3.2 & \,3.2 & \,2.8 & \,2.9 & \,2.8 & \,2.9 & \,2.9 & \,2.3 & \,2.2 & \,2.2 & \,2.3 & \,2.4\\
    & & \ellipse & \,4.8 & \,3.1 & \,2.5 & \,3.8 & \,4.9 & \,4.9 & \,3.8 & \,4.3 & \,4.0 & \,4.4 & \,4.9 & \,3.0 & \,3.5 & \,3.7 & \,3.6 & \,3.4\\
    & & \vvflows & \textbf{\,2.2} & \textbf{\,1.9} & \,1.8 & \textbf{\,2.1} & \textbf{\,2.5} & \textbf{\,2.4} & \textbf{\,2.0} & \textbf{\,2.2} & \textbf{\,2.1} & \textbf{\,2.0} & \textbf{\,2.0} & \textbf{\,1.7} & \textbf{\,1.9} & \textbf{\,1.9} & \textbf{\,1.7} & \textbf{\,1.6}\\
    & & \vvflowspy & \,2.3 & \textbf{\,1.9} & \,1.9 & \,2.2 & \textbf{\,2.5} & \textbf{\,2.4} & \textbf{\,2.0} & \,2.3 & \,2.2 & \,2.1 & \textbf{\,2.0} & \textbf{\,1.7} & \,2.0 & \,2.0 & \textbf{\,1.7} & \textbf{\,1.6}\\

    \cmidrule{2-19}
      & \multirow{4}{*}{\dphill}
      & \vweight & \,2.0 & \,2.0 & \,1.5 & \,1.4 & \,1.9 & \,2.1 & \,2.0 & \,1.9 & \,1.9 & \,2.0 & \,2.2 & \,2.0 & \,1.5 & \,1.6 & \,1.9 & \,2.0\\
    & & \ellipse & \,2.2 & \,2.1 & \,1.5 & \textbf{\,1.3} & \,1.8 & \,2.3 & \,2.4 & \,2.3 & \,2.3 & \,2.5 & \,2.8 & \,2.4 & \,1.7 & \,1.9 & \,2.4 & \,2.6\\
    & & \vvflows & \textbf{\,1.5} & \textbf{\,1.5} & \textbf{\,1.4} & \,1.4 & \textbf{\,1.6} & \textbf{\,1.6} & \textbf{\,1.5} & \textbf{\,1.5} & \textbf{\,1.5} & \textbf{\,1.5} & \textbf{\,1.6} & \textbf{\,1.5} & \textbf{\,1.3} & \textbf{\,1.4} & \textbf{\,1.5} & \textbf{\,1.5}\\
    & & \vvflowspy & \textbf{\,1.5} & \textbf{\,1.5} & \textbf{\,1.4} & \,1.4 & \textbf{\,1.6} & \textbf{\,1.6} & \textbf{\,1.5} & \textbf{\,1.5} & \textbf{\,1.5} & \textbf{\,1.5} & \textbf{\,1.6} & \,1.5 & \textbf{\,1.3} & \textbf{\,1.4} & \textbf{\,1.5} & \textbf{\,1.5}\\

    \bottomrule
    \end{tabular}
\end{table*}

\begin{table*}[tp]
    \centering
    \caption{Efficiency for finding a solution in each bin of the respective unfolded double differential distributions with \vweight and \ellipse.
    The bins are ordered first by increasing \mttbar followed by the second variable, with vertical dividers indicating the bin edges in \mttbar.
    The efficiency of \vvflows in all bins is 100\%.
    }
    \label{tab:app:sol_efficiency}
    \begin{tabular}{c c r r r r r | r r r r | r r r r | r r r r}
    \toprule
    \multicolumn{2}{c}{Observables} & Method & \multicolumn{16}{c}{Efficiency at finding a solution (per bin) [\%]}\\
    \midrule

    \multirow{8}{*}{\wrapentry{0.07\textwidth}{\mttbar}} & \multirow{2}{*}{\ytt}
    & \vweight & \, 97 & \, 98 & \, 97 & \, 97 & \, 96 & \, 96 & \, 97 & \, 96 & \, 93 & \, 94 & \, 94 & \, 94 & \, 89 & \, 90 & \, 89 & \, 89 \\
    & & \ellipse & \, 91 & \, 92 & \, 92 & \, 91 & \, 84 & \, 82 & \, 84 & \, 82 & \, 67 & \, 68 & \, 68 & \, 70 & \, 58 & \, 54 & \, 51 & \, 50 \\

    \cmidrule{2-19}
      & \multirow{2}{*}{\pttt}
      & \vweight & \, 99 & \, 95 & \, 91 & \, 89 & \, 98 & \, 95 & \, 90 & \, 84 & \, 97 & \, 92 & \, 89 & \, 82 & \, 94 & \, 88 & \, 81 & \, 81 \\
      & & \ellipse & \, 96 & \, 86 & \, 74 & \, 59 & \, 88 & \, 79 & \, 66 & \, 50 & \, 75 & \, 66 & \, 52 & \, 38 & \, 60 & \, 52 & \, 34 & \, 37 \\

    \cmidrule{2-19}
      & \multirow{2}{*}{\pttop}
      & \vweight & \, 98 & \, 96 & \, 92 & \, 86 & \, 97 & \, 97 & \, 96 & \, 91 & \, 95 & \, 95 & \, 95 & \, 93 & \, 96 & \, 89 & \, 90 & \, 89 \\
    & & \ellipse & \, 95 & \, 88 & \, 72 & \, 51 & \, 89 & \, 86 & \, 80 & \, 63 & \, 77 & \, 74 & \, 72 & \, 63 & \, 61 & \, 56 & \, 55 & \, 51 \\

    \cmidrule{2-19}
      & \multirow{2}{*}{\dphill}
      & \vweight & \, 96 & \, 97 & \, 98 & \, 99 & \, 96 & \, 96 & \, 97 & \, 97 & \, 93 & \, 94 & \, 94 & \, 94 & \, 90 & \, 90 & \, 91 & \, 88 \\
    & & \ellipse & \, 90 & \, 92 & \, 92 & \, 94 & \, 84 & \, 84 & \, 83 & \, 82 & \, 76 & \, 74 & \, 68 & \, 64 & \, 57 & \, 57 & \, 57 & \, 48 \\

    \bottomrule
    \end{tabular}
\end{table*}
  
  \subsection{Direct regression}

\begin{figure*}[htb]
    \includegraphics[width=0.3\textwidth]{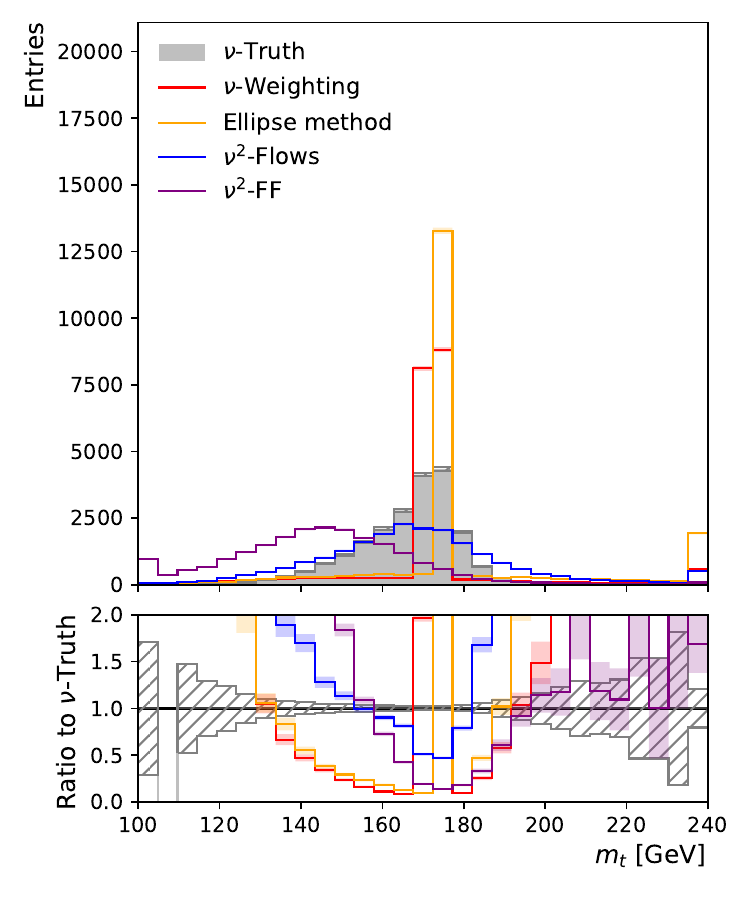}
    \includegraphics[width=0.3\textwidth]{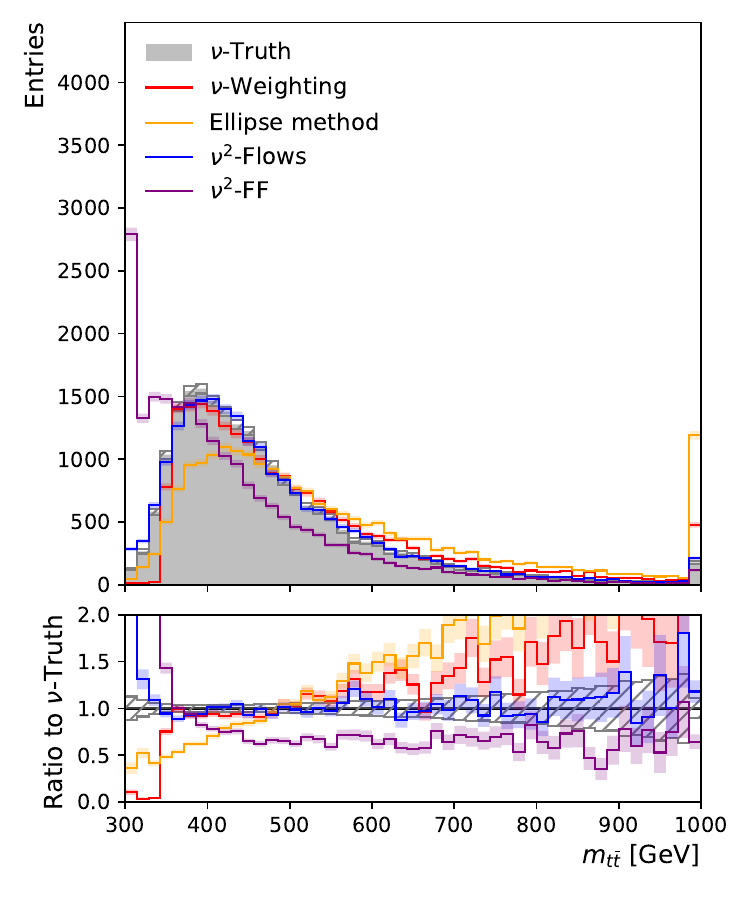}
    \caption{The invariant masses of the reconstructed top quark and $t\bar t$ system when using the three neutrino reconstruction methods discussed in the paper, as well as a feed-forward regression model $\nu^2$-FF, in comparison to \vtruth (shaded grey). This plot highlights large negative bias induced by the feed-forward model.
    }
    \label{fig:app:nu_ff}
\end{figure*}

Strong biases are introduced when replacing the normalizing flow with a na\"ive feed-forward network to predict the neutrino momenta for each event directly.
This network, $\nu^2$-FF, has the same architecture as the feature extraction network in \vvflows, but with a dense output layer for the neutrino momenta, and is trained with an L2 loss.
\Cref{fig:app:nu_ff} shows the reconstructed invariant mass of the top quark and \ttbar pair when using $\nu^2$-FF to reconstruct the neutrinos in comparison to the other reconstruction approaches.
A strong negative bias can be seen in both distributions.

  \subsection{Origin of improvement over \vweight}

In the \vweight a weight is used to find the best solution, as defined in \cref{eq:neutrino_weight}, which can also be interpreted as how good a solution is.
The improvement from \vvflows could arise from the events with low values of $w$ or no solutions, or come from all events regardless of how well \vweight performs.

\begin{figure*}
    \includegraphics[width=0.32\textwidth]{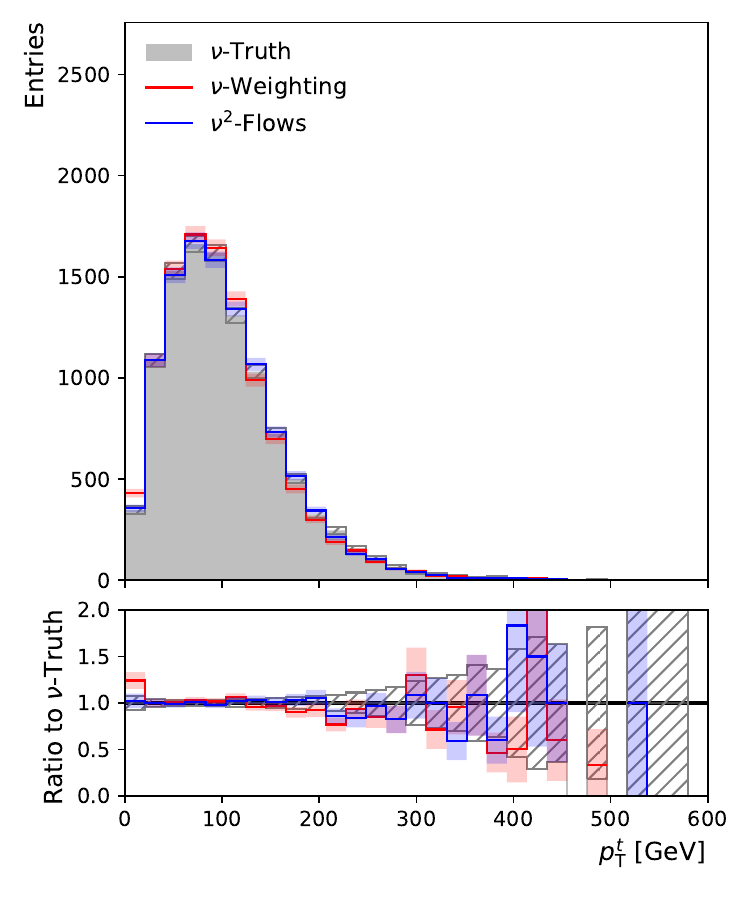}
    \includegraphics[width=0.32\textwidth]{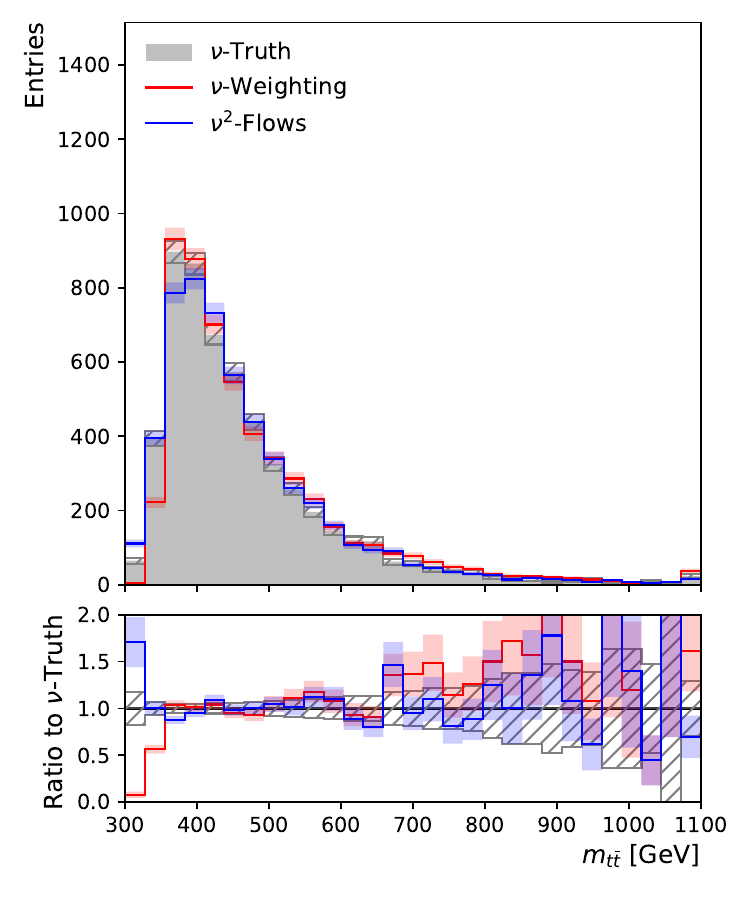}
    \includegraphics[width=0.32\textwidth]{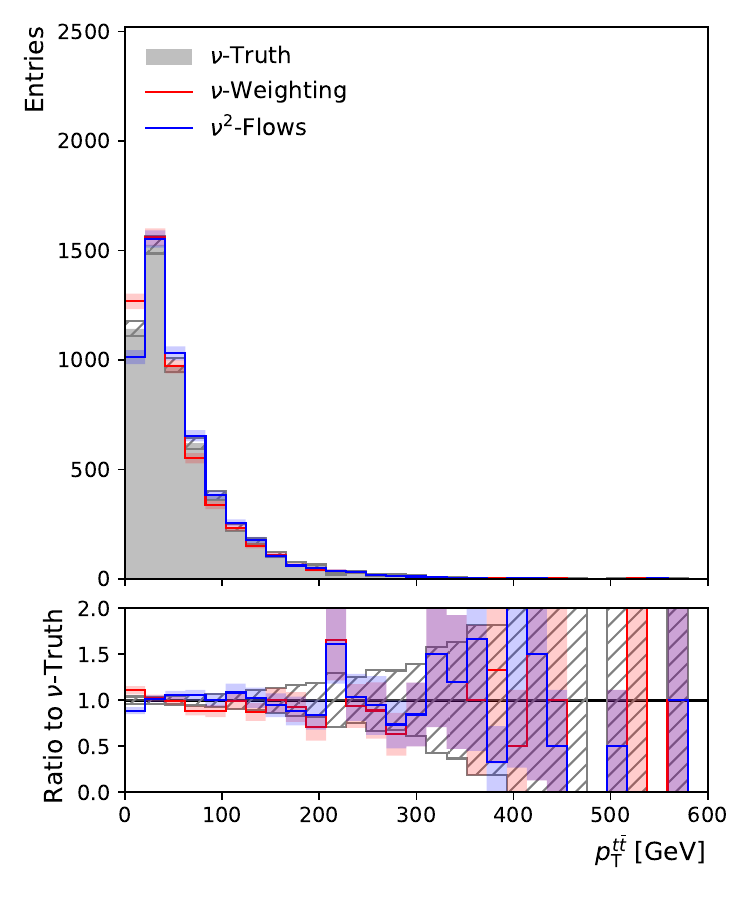}
    \includegraphics[width=0.32\textwidth]{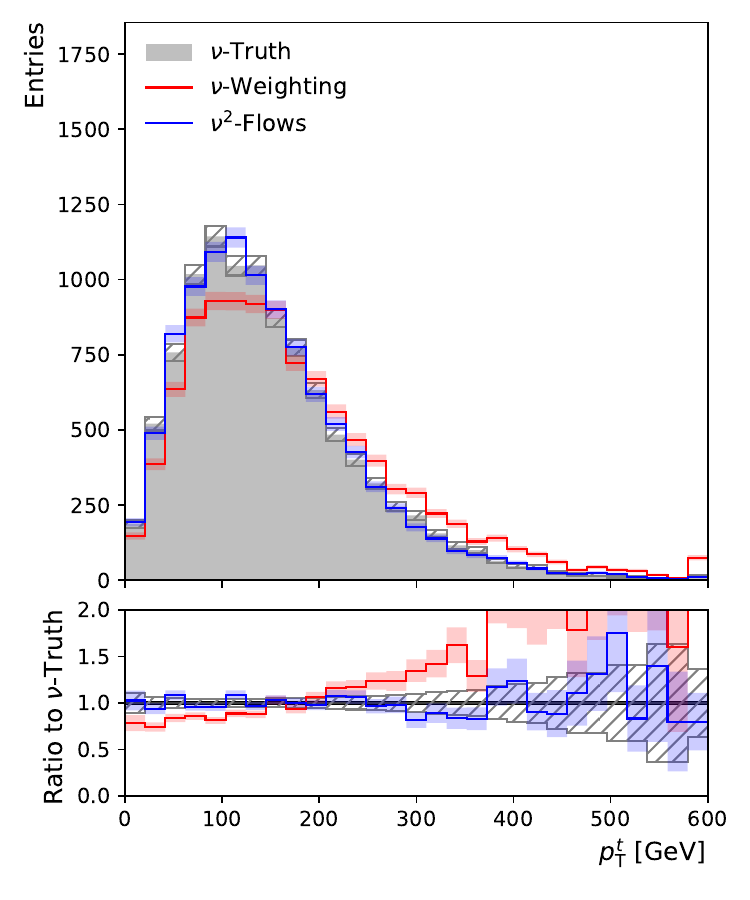}
    \includegraphics[width=0.32\textwidth]{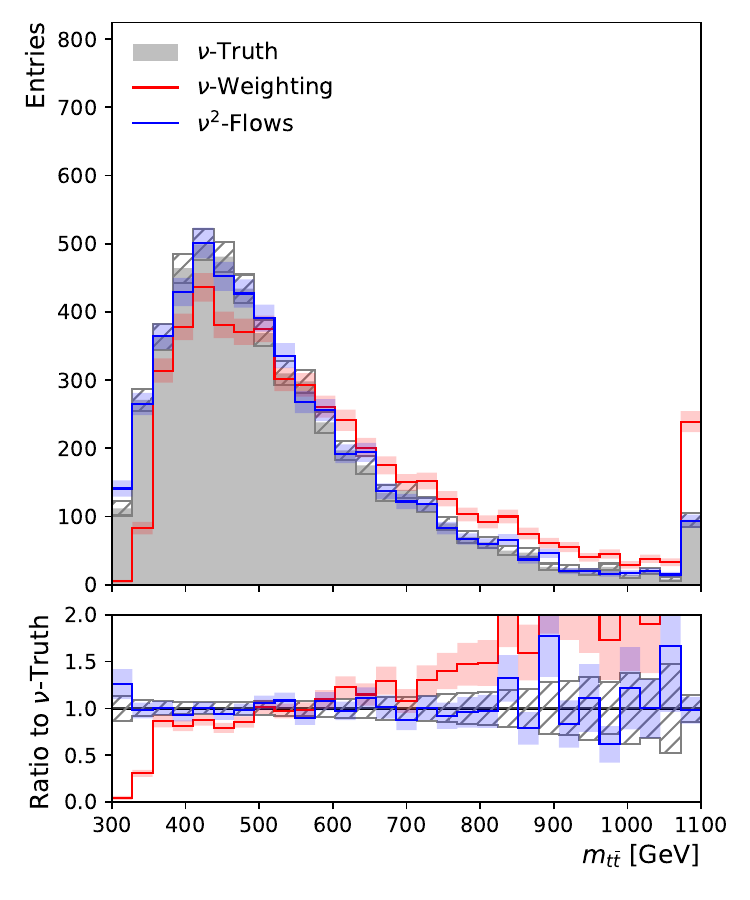}
    \includegraphics[width=0.32\textwidth]{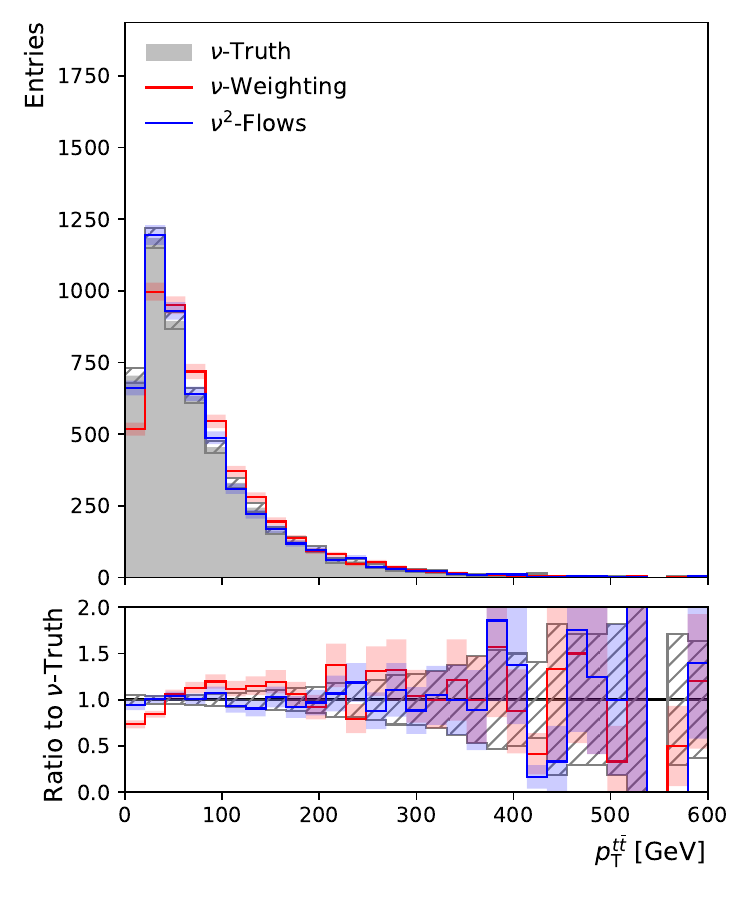}
    \caption{The reconstructed top quark \pt, and the invariant mass and \pt of the reconstructed \ttbar system when using \vvflows and \vweight in comparison to the two baseline approaches and \vtruth (shaded grey) for events where \vweight returns $w > 0.95$ (top) and $w < 0.8$ (bottom).}
    \label{fig:app:vweightbad}
\end{figure*}

To investigate this we compare the reconstruction performance for events where \vweight has either a good solution or a poor solution.
We define good performance as values of $w > 0.9$ and poor performance as $w < 0.3$.
The \pttop, \pttt and \mttbar distributions with these selections are shown in \cref{fig:app:vweightbad}.
Although substantial improvement can be seen over \vweight in regions where \vweight has a low weight, \vvflows still exhibits improved agreement with \vtruth.
It is equally encouraging that there is no noticeable difference in the quality of reconstruction with \vvflows for events that can be considered well reconstructed or poorly reconstructed by \vweight.

  \subsection{Additional figures}

\begin{figure*}[htbp]
    \centering
    \includegraphics[width=0.32\textwidth]{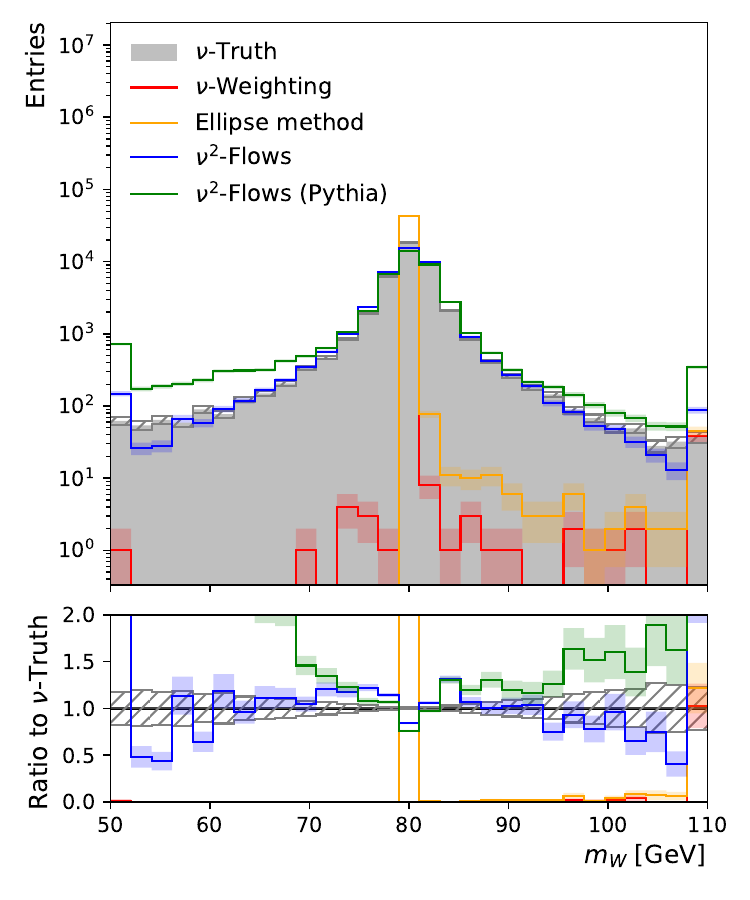}
    \includegraphics[width=0.32\textwidth]{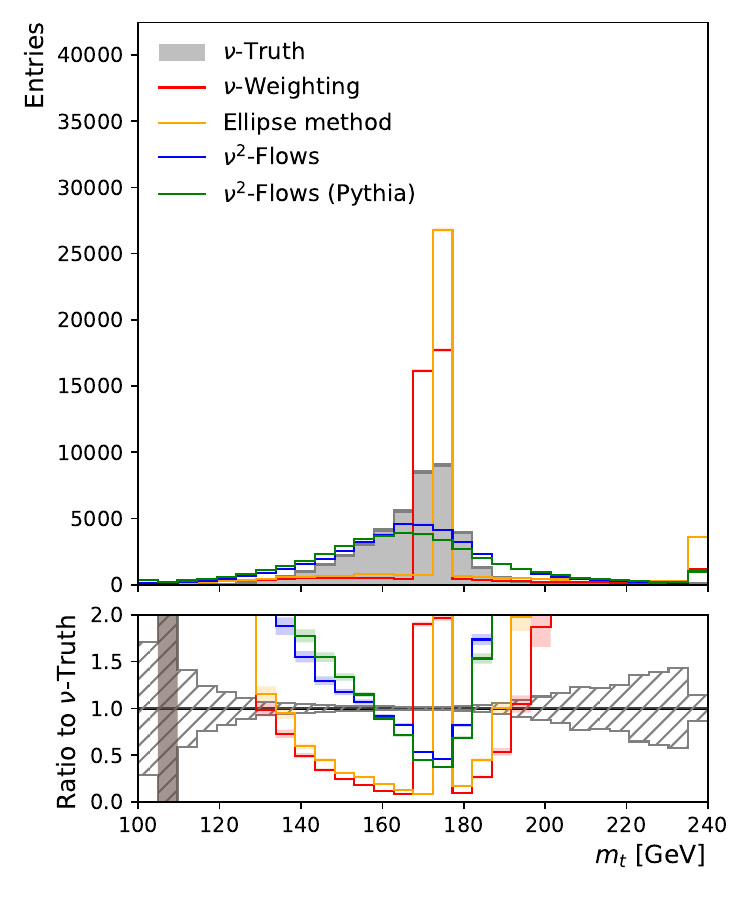}
    \caption{The reconstructed invariant mass of the \Wboson and top quarks when using the three neutrino reconstruction methods in comparison to \vtruth (shaded grey) as well as the alternative \vvflowsPy model.
    }
    \label{fig:pythiaW}
\end{figure*}

\begin{figure*}
    \centering
    \includegraphics[width=0.32\textwidth]{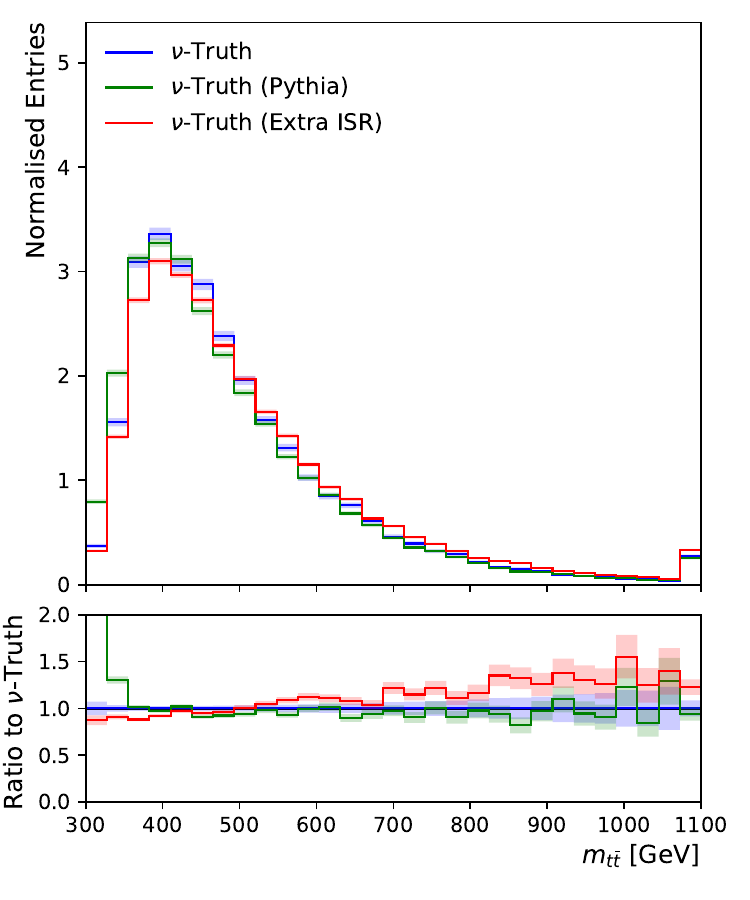}
    \includegraphics[width=0.32\textwidth]{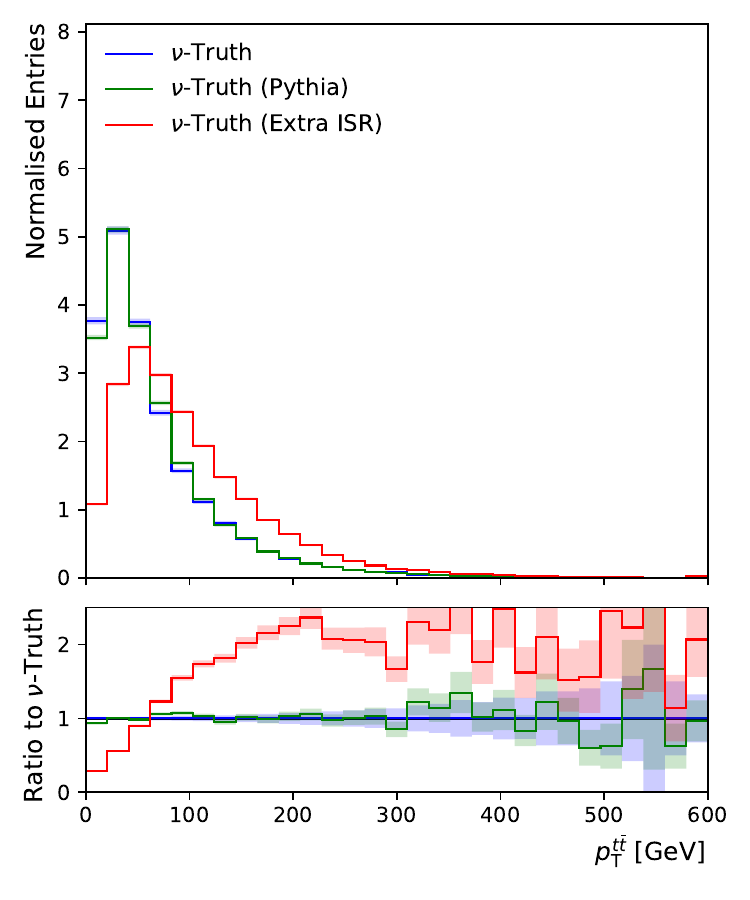}
    \includegraphics[width=0.32\textwidth]{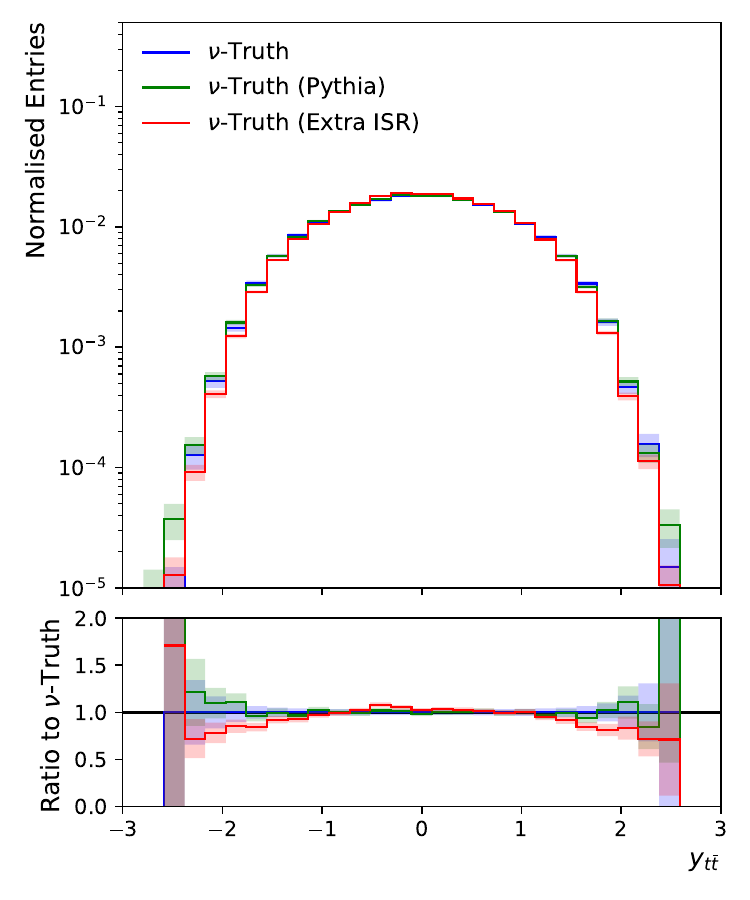}\\

    \includegraphics[width=0.32\textwidth]{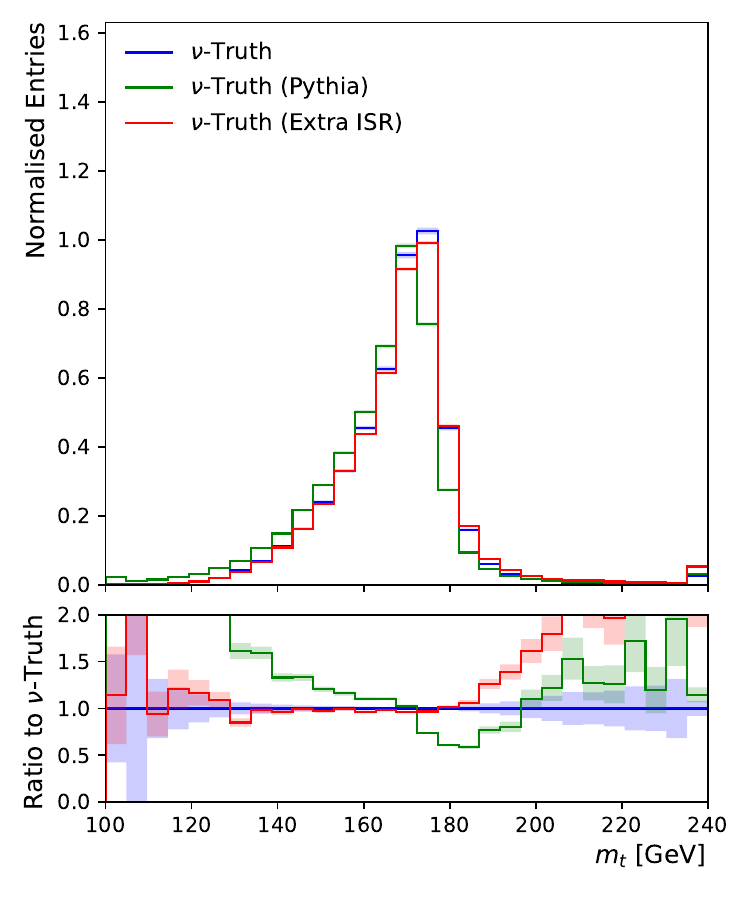}
    \includegraphics[width=0.32\textwidth]{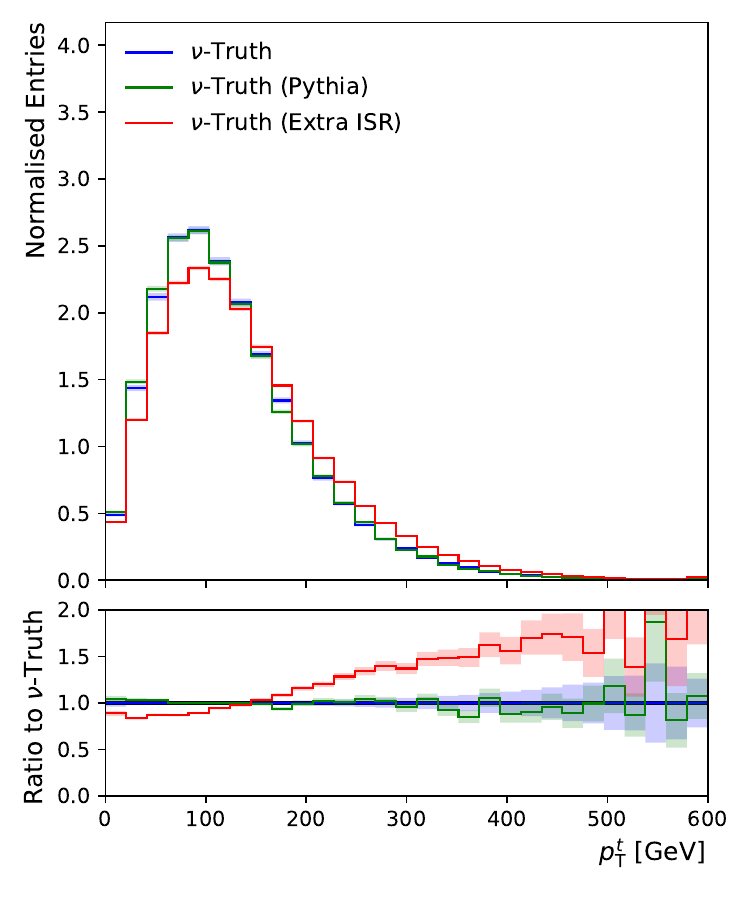}
    \caption{The invariant mass, \pt, and rapidity of the reconstructed \ttbar system (top row) and the invariant mass, and \pt of the reconstructed top quarks (bottom row) for \vtruth with the three independent simulated samples.
    }
    \label{fig:truthdiffs_t}
\end{figure*}

\begin{figure*}
    \centering
    \includegraphics[width=0.32\textwidth]{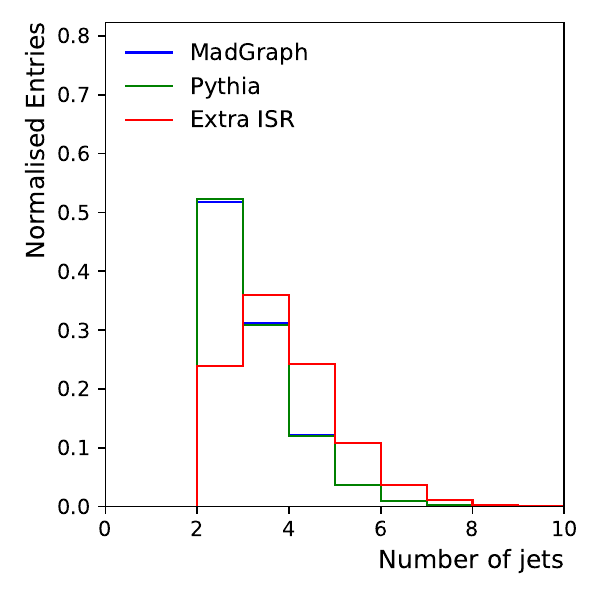}
    \includegraphics[width=0.32\textwidth]{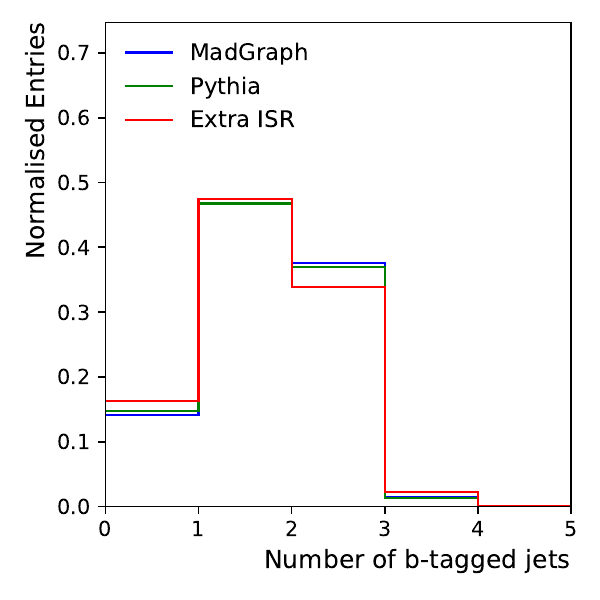}
    \caption{The jet and $b$-jet multiplicities of the three independent simulated samples.
    }
    \label{fig:truthdiffs_multi}
\end{figure*}

The reconstructed top quark and \Wboson invariant masses are compared in \cref{fig:pythiaW}.
The \vtruth distributions for the nominal, alternative and extra ISR sample are compared in \cref{fig:truthdiffs_t}, and the jet and $b$-jet multiplicities for the three samples are compared in \cref{fig:truthdiffs_multi}.

\begin{figure*}[htp]
    \includegraphics[width=0.32\textwidth]{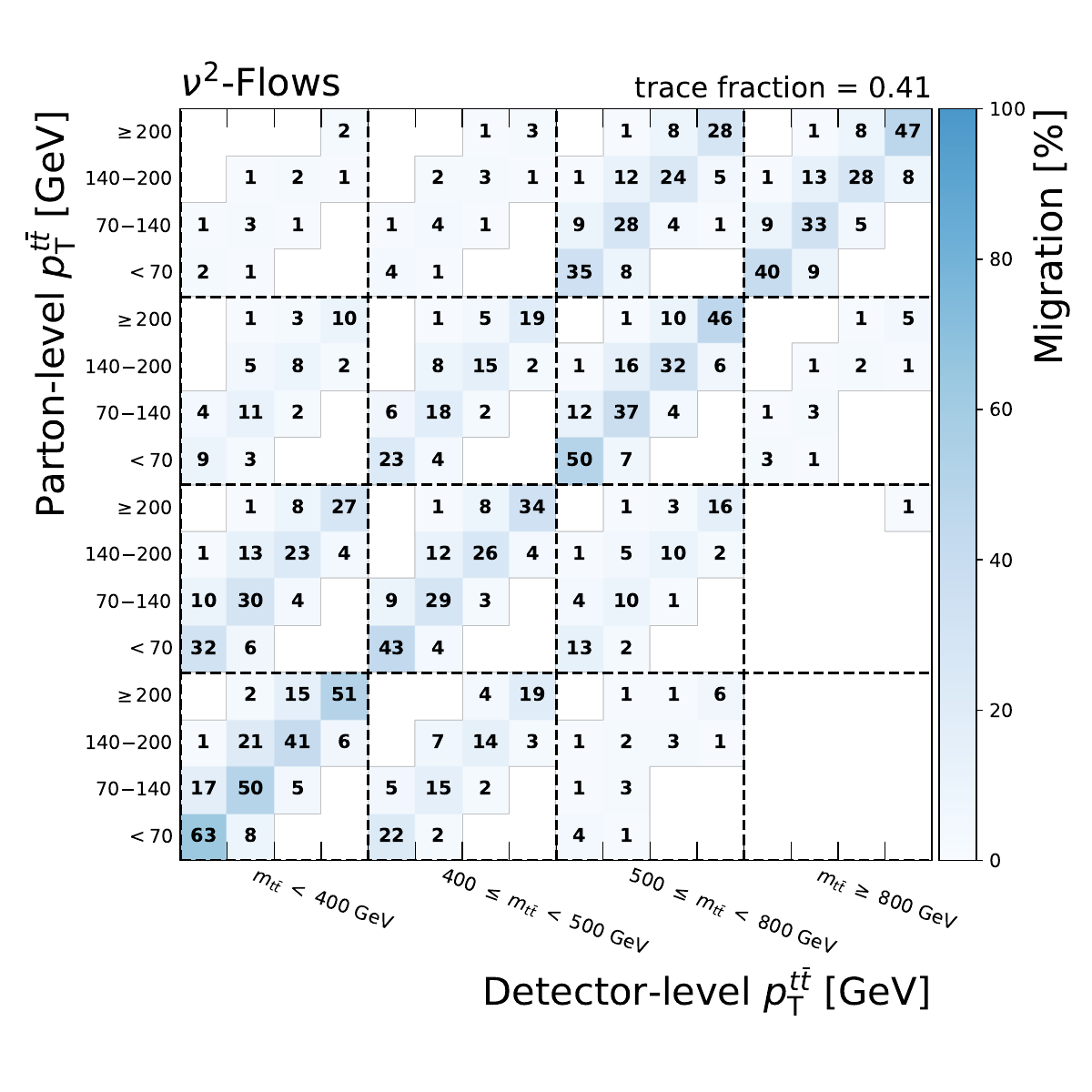}
    \includegraphics[width=0.32\textwidth]{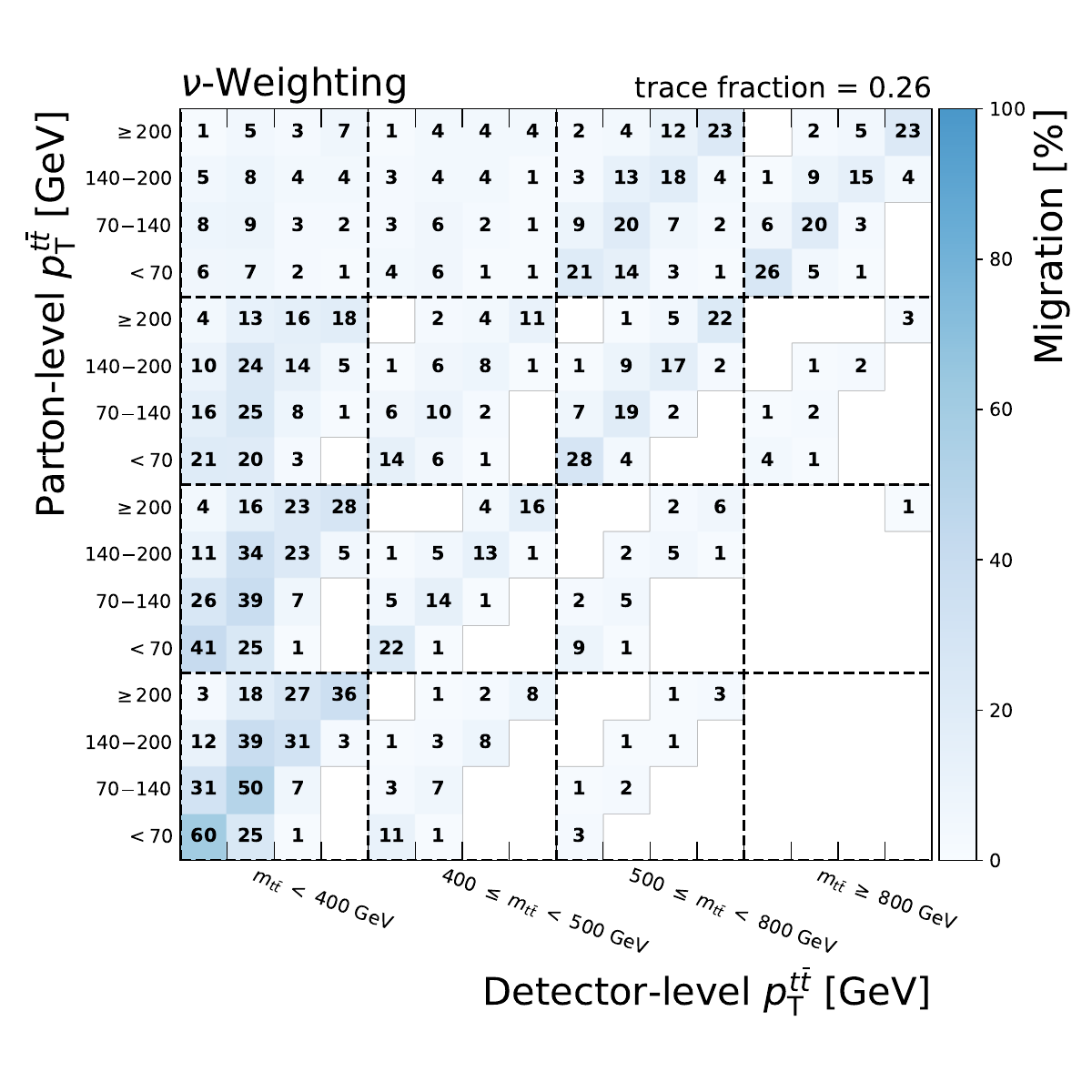}
    \includegraphics[width=0.32\textwidth]{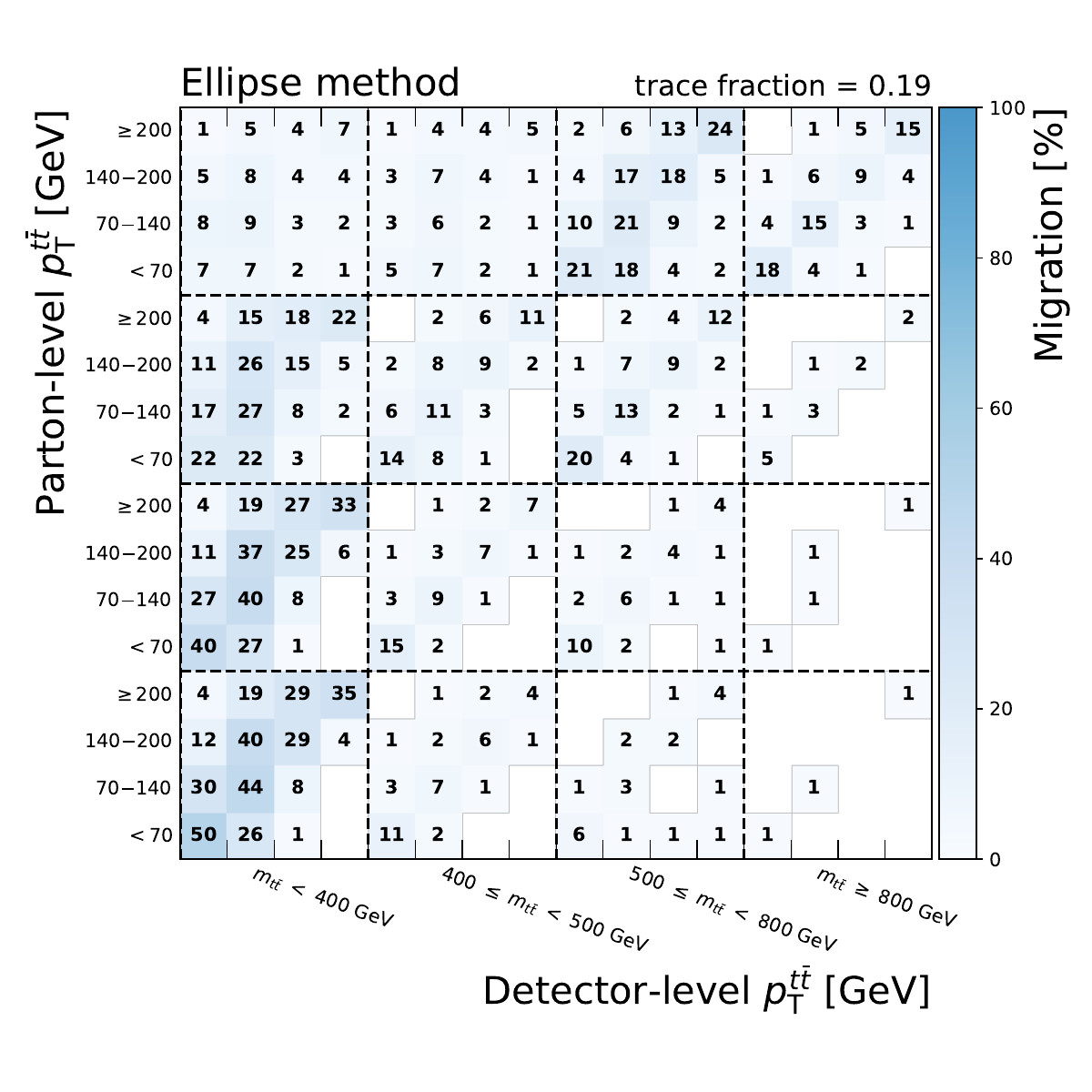}

    \includegraphics[width=0.32\textwidth]{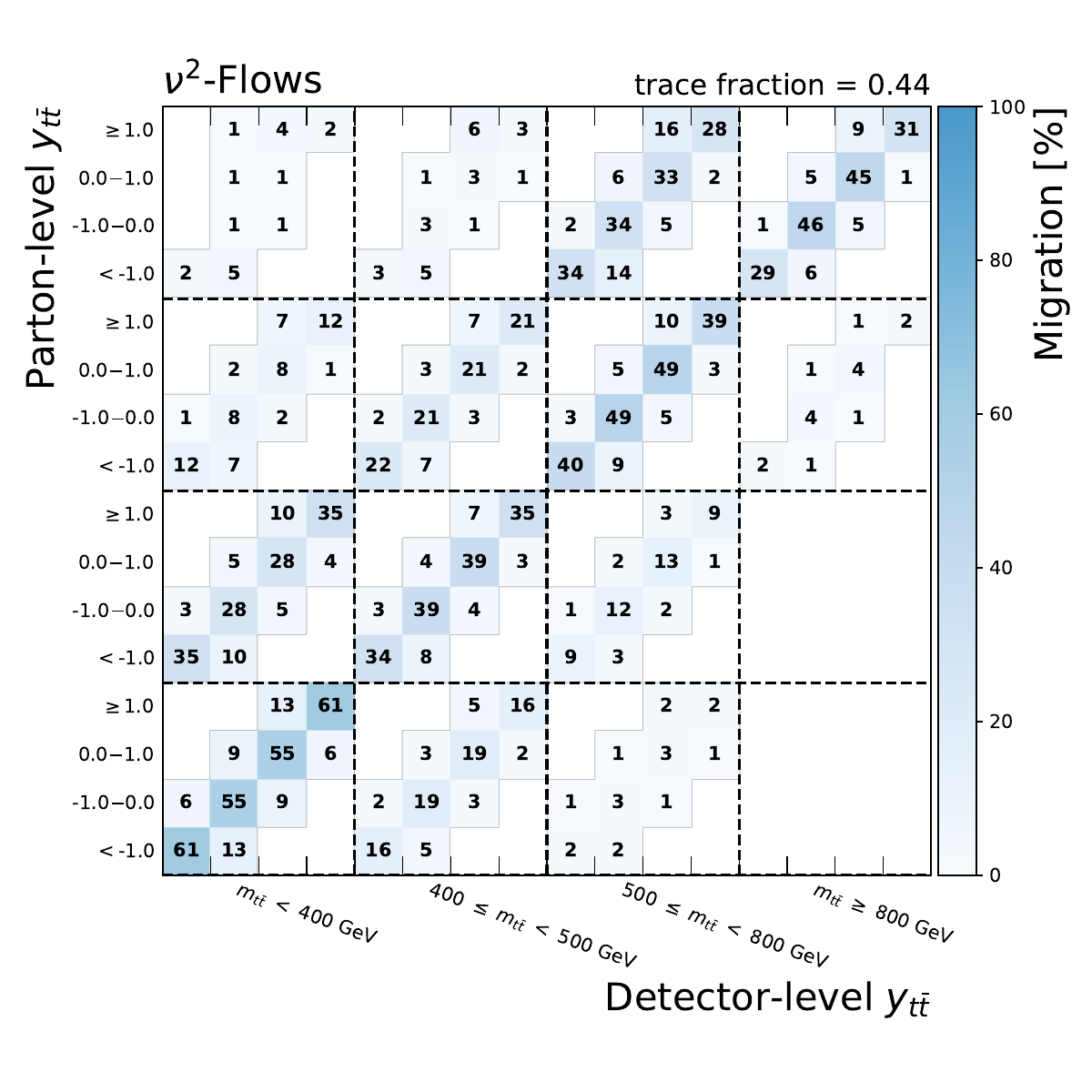}
    \includegraphics[width=0.32\textwidth]{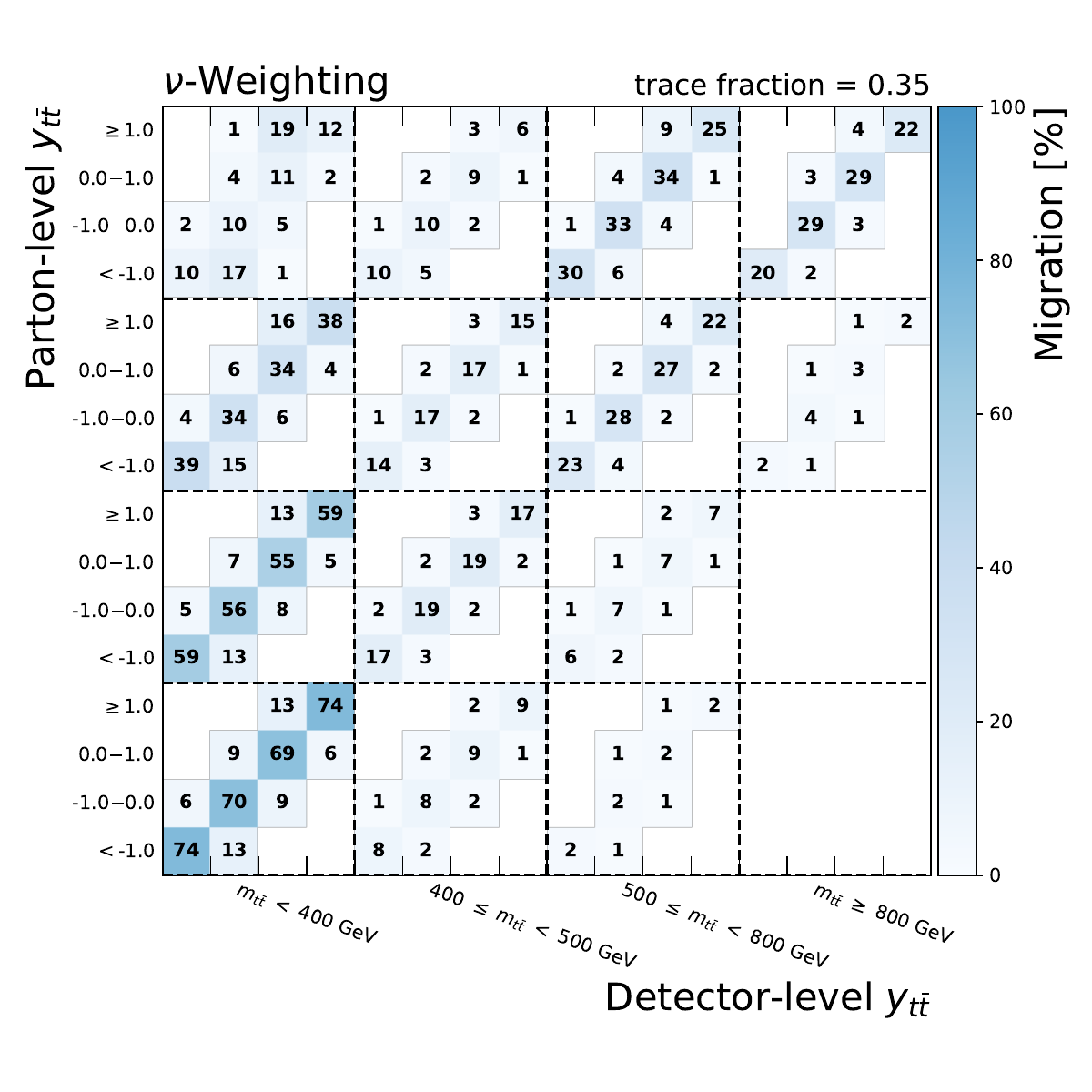}
    \includegraphics[width=0.32\textwidth]{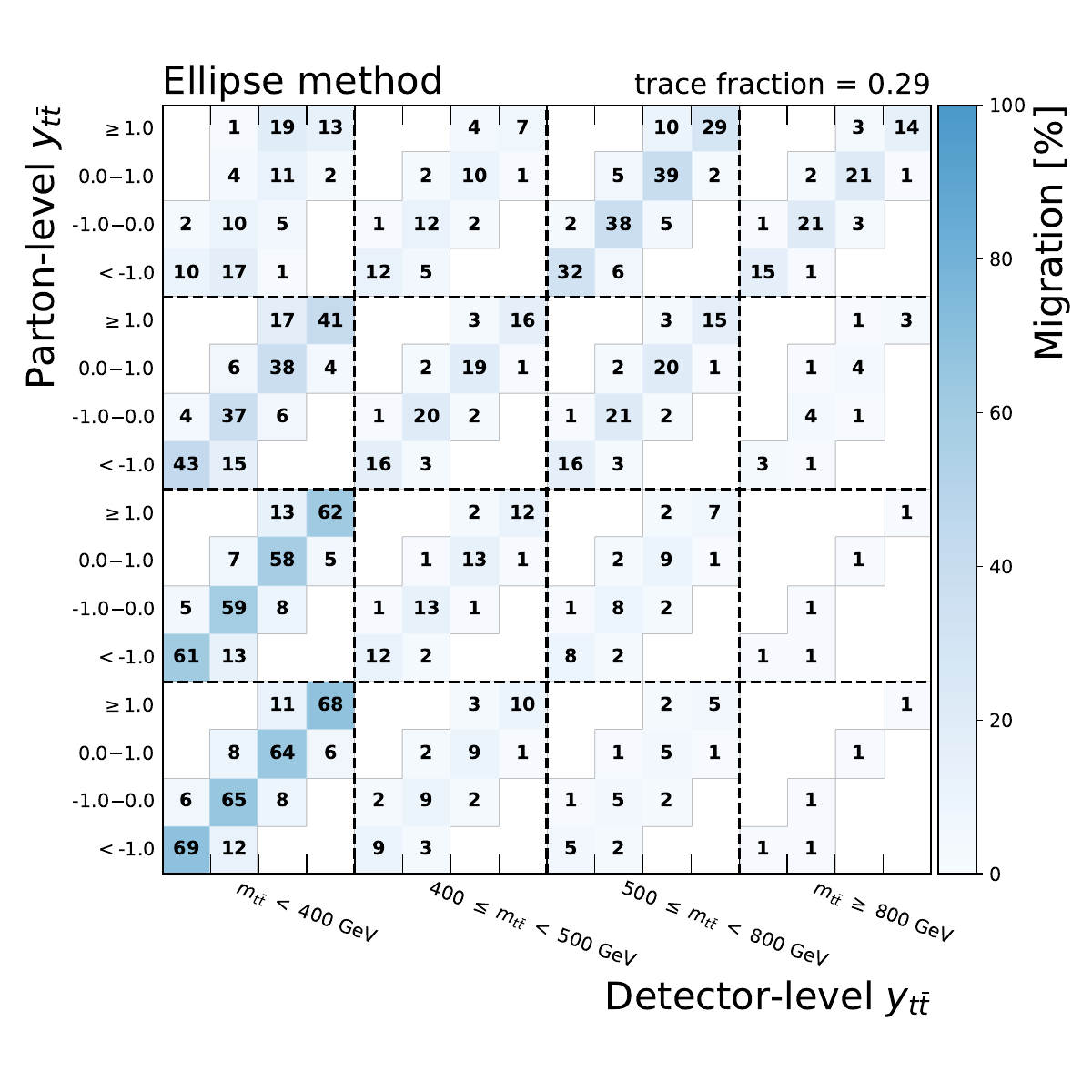}
    \caption{Binned response matrices for the double differential measurement of  \mttbar and \pttt (top) and \mttbar and \ytt (bottom) when using each of the three methods for neutrino reconstruction. The binning is symmetric for both the parton and detector level observables, however the \mttbar bins are labelled on the $x$-axis with the \ytt bins labelled on the $y$-axis.
    The trace fraction is calculated for each method for a simple quantitative comparison and is 0.70 when using \vtruth.}
    \label{fig:unfold_ytt}
\end{figure*}

  The response matrices for the double differential distributions in \mttbar and \pttop, and \mttbar and \ytt, are shown in \cref{fig:unfold_ytt}.

\begin{figure*}[htb]
    \centering
    \includegraphics[width=0.3\textwidth,trim=0 70 45 0, clip]{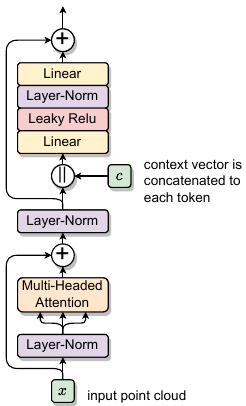}
    \hskip4ex
    \includegraphics[width=0.43\textwidth,trim=0 00 0 0, clip]{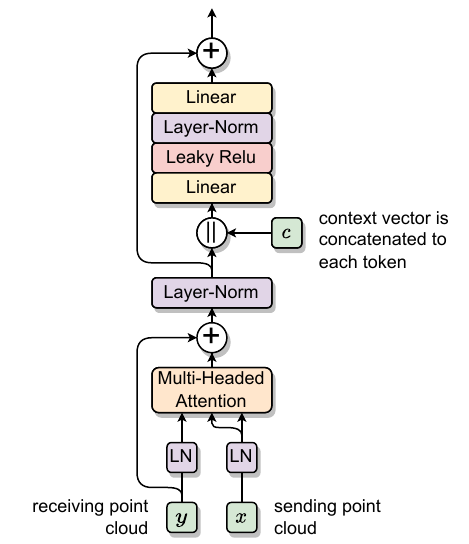}
    \caption{Transformer encoder block (left) and cross-attention block (right) comprising multi-headed attention, layer normalisation and simple linear layers. 
    Residual connections are used after the multi-headed attention and linear operation.
    Conditional information is provided as context by concatenating it to each token before the linear layer.}
    \label{fig:te_block}
\end{figure*}

  Detailed schema of the transformer encoder block and cross-attention block used in \vvflows are detailed in \cref{fig:te_block}. 

\end{document}